\documentclass[twocolumn,times]{aastex63}

\usepackage{natbib}

\newcommand{\ten}[2]{$#1$$\times10^{#2}$}
\newcommand{\isotope}[2]{\ensuremath{\mathrm {^{#2}#1}}}
\newcommand{\xnet}{{XNet}}
\newcommand{\chimera}{{\sc Chimera}}
\newcommand{\kepler}{{\sc Kepler}}

\newcommand{\vertex}{{\sc Vertex-Prometheus}}
\newcommand{\hotb}{{\sc Prometheus-HOTB}}
\newcommand{\msun}{\ensuremath{M_\sun}}
\newcommand{\tmap}{\ensuremath{t_{\rm map}}}
\newcommand{\spun}{{\sc 2d3d}}
\newcommand{\tilt}{{\sc 2d3d$_{\rm Tilted}$}}
\newcommand{\threed}{{\sc 3d3d}}
\newcommand{\twod}{{\sc 2d2d}}
\newcommand{\kms}{\ensuremath{\rm km \; s^{-1}}}

\newcommand{\UTphys}{Department of Physics and Astronomy, University of Tennessee, Knoxville, TN 37996-1200, USA}
\newcommand{\ORNLphys}{Physics Division, Oak Ridge National Laboratory, P.O. Box 2008, Oak Ridge, TN 37831-6354, USA}
\newcommand{\NCCS}{National Center for Computational Sciences, Oak Ridge National Laboratory, P.O. Box 2008, Oak Ridge, TN 37831-6164, USA}

\begin{document}
	\title{Three Dimensional Core-Collapse Supernova Simulations with 160 Isotopic Species Evolved to Shock Breakout}
	
	\author[0000-0002-5088-4487]{Michael A. Sandoval}
	\affiliation{\UTphys}
	\affiliation{\ORNLphys}
	\affiliation{\NCCS}

	\author[0000-0002-9481-9126]{W. Raphael Hix}
	\affiliation{\ORNLphys}
	\affiliation{\UTphys}

	\author[0000-0002-5358-5415]{O. E. Bronson Messer}
	\affiliation{\NCCS}
	\affiliation{\ORNLphys}
	\affiliation{\UTphys}

	\author[0000-0002-5231-0532]{Eric J. Lentz}
	\affiliation{\UTphys}
	\affiliation{\ORNLphys}

	\author[0000-0003-3023-7140]{J. Austin Harris}
	\affiliation{\NCCS}

	\shorttitle{3D CCSN Breakout}
	\shortauthors{Sandoval et al.}
	
	\accepted{August 11, 2021}
	\published{November 8, 2021}
	\submitjournal{ApJ}
	
	\begin{abstract}
		We present three-dimensional simulations of core-collapse supernovae using the FLASH code that follow the progression of the explosion to the stellar surface, starting from neutrino-radiation hydrodynamic simulations of the neutrino-driven phase performed with the \chimera\ code.
		We consider a 9.6-\msun\ zero-metallicity progenitor starting from both 2D and 3D \chimera\ models, and a 10-\msun\ solar-metallicity progenitor starting from a 2D \chimera\ model, all simulated until shock breakout in 3D while tracking 160 nuclear species.
		The relative velocity difference between the supernova shock and the metal-rich Rayleigh-Taylor (R-T) ``bullets'' determines how the metal-rich ejecta evolves as it propagates through the density profile of the progenitor and dictates the final morphology of the explosion.
		We find maximum \isotope{Ni}{56} velocities of $\sim$1950~\kms\ and $\sim$1750~\kms\ at shock breakout from 2D and 3D 9.6-\msun\ \chimera\ models, respectively, due to the bullets' ability to penetrate the He/H shell.
		When mapping from 2D, we find that the development of higher velocity structures is suppressed when the 2D \chimera\ model and 3D FLASH model meshes are aligned.
		The development of faster growing spherical-bubble structures, as opposed to the slower growing toroidal structure imposed by axisymmetry, allows for interaction of the bullets with the shock and seeds further R-T instabilities at the He/H interface.
		We see similar effects in the 10-\msun\ model, which achieves maximum \isotope{Ni}{56} velocities of $\sim$2500~\kms\ at shock breakout.
	\end{abstract}
	
	
	\section{Introduction} \label{intro}
	It has been clear from the earliest observations of supernova remnants that large-scale asymmetries develop in the decades between the explosion and the present day.
	Modern observations continue to reveal more detail.
	Direct imaging of \isotope{Ti}{44} emission in Cassiopeia~A \citep{GrHaBo14,GrFrHa17} revealed previously hidden asymmetries in the innermost ejecta.
	Observations of \isotope{Ti}{44} ejection velocities in SN~1987A \citep{BoHaMi15} suggest an even higher level of asymmetry in that supernova.
	X-ray observations of G292.0+1.8 \citep{BhPaSc19} reveal gross elemental asymmetries in the ejecta of this young, oxygen-rich, Galactic supernova remnant, echoing earlier work on Cassiopeia~A \citep{HuRaBu00}.
	In this work, we present state-of-the-art core-collapse supernova (CCSN) simulations that explore the development and evolution of these asymmetries as the supernova shock progresses through the entire star. 
	
	Observations at earlier epochs support the assertion that CCSN explosions are asymmetric from their earliest days \citep{ArBaKi89,McCr93,Wood97,Mull98}.
	Not surprisingly, evidence from the closest supernova in modern times, SN~1987A, is particularly extensive \citep{WaWhHo02,LaFrSp16}.
	Observed asymmetries in iron lines have been explained by the concentration of iron-peak elements into high-velocity ``bullets'' \citep{SpMeAl90}.
	Similar bullets have been invoked to explain features of the Vela supernova remnant \citep{AsEgTr95, StJoVe95}.
	The early development of fine structure in the $H_{\alpha}$ line in SN~1987A, less than a month after the explosion, \citep[referred to as the \emph{Bochum event},][]{HaThDa88}, was explained by \citet{UtChAn95} as the result of a large ($\sim$10$^{-3}$~\msun) clump of nickel ejected at high velocity ($\approx$4,700~\kms) into the far hemisphere of the supernova.
	Near-IR observations of \ion{He}{1} lines arising roughly two months after the explosion of SN~1987A were similarly interpreted as indications of dense clumps of \isotope{Ni}{56} mixed into the hydrogen envelope \citep{FaMe99}.
	Subsequent observations \citep{SiWeRe13} from different viewing angles via light echo spectroscopy support a strongly asymmetric distribution of nickel.
	
	Evidence for asymmetries in SN~1987A set in motion the earliest multidimensional studies of supernova shock propagation \citep[see, e.g.,][]{HaMaNo90,MuFrAr91,HeBe92}.
	These studies revealed that the supernova shock's encounters with the stellar compositional interfaces induced Rayleigh-Taylor (R-T) instabilities that effectively broke spherical symmetry.
	Specifically, it has been shown that R-T instabilities originate at the Si/O, (C+O)/He, and He/H boundaries of the star, and that these instabilities can shape the ejecta \citep{Chev76,HaMaNo90,FrMuAr91,MuFrAr91,HeBe92,NaShSa98,KaArRe00,JoAlWo10}.
	However, the asymmetry introduced was not sufficient to explain the observed asymmetries in SN~1987A, suggesting asymmetries are part of the central engine of the explosion, leading to the earliest multidimensional investigations of that central engine \citep{MiWiMa93,HeBeHi94,BuHaFr95,JaMu96}.
	
	These studies show us that the large-scale features associated with the explosion are directly tied to the asymmetries formed at early times due to the explosion mechanism itself.
	As the core collapses, a strong shock is unleashed into the surrounding composition layers of the progenitor.
	This shock eventually stalls due to nuclear dissociation and loss of energy in the form of neutrinos.
	Although the shock stalls for a hundred milliseconds or more, the explosion is eventually able to continue due to neutrino heating above the proto-neutron star (PNS) formed after collapse.
	As the explosion develops further, large plumes begin to form and dominate the morphology of the system \citep[see, e.g., ][]{MeJaBo15,LeBrHi15,VaBuRa19}.
	Although developed within the inner 1000~km, these plumes seed further asymmetries as the explosion progresses, and echoes of them can be seen hours later at the surface of the star ($\sim 10^8$~km).
	These plumes typically align with the poles (axis of symmetry) in 2D and exhibit random alignment in 3D.
	However, there is a significant gap between where current three-dimensional supernova simulations cease and where the ejecta could potentially be seen, which limits our ability to compare to observations.
	
	\citet{KiPlJa03} extended neutrino-driven multi-dimensional CCSN simulations to shock breakout in 2D while analyzing $^{56}$Ni clump formation along the way.
	Previously, most late-time explosion simulations were initiated with parameterized spherical pistons or thermal bombs rather than a neutrino heating simulation.
	Although in axisymmetry, the work of \citet{KiPlJa03} represented a more faithful attempt at understanding the generation and propagation of $^{56}$Ni bullets through the star, which  at the time, displayed a  discrepancy between observed and simulated velocities \citep{HeBe92}.
	\citet{KiPlJa03} discovered that their 2D models displayed significant differences in the ejecta when compared to previous piston initiated simulations that did not accurately capture the growth of the R-T instabilities.
	This motivated further exploration of the crucial impact of the stellar density structure on the evolution of the bullets prior to shock breakout.
	
	\citet{HaJaMu10} explored these issues using a series of 2D and 3D shock breakout models powered by neutrino heating.
	They found that the R-T instabilities generated were different than those discussed in simpler 3D simulations \citep{NaNaMi88,MuHiOr89,YaNaOo90}, and that the propagation of bullets in 3D behaved differently than in 2D.
	In agreement with \citet{KaArRe00}, they showed that the inherent axisymmetry of 2D models leads to slower clumps compared to those in 3D, due to enhanced kinematic drag relative to the buoyant force.
	In this case, a 2D model has toroidal structures due to axisymmetry, whereas a 3D model has bubble structures that are more spherical.
	The density profile of the star determines how unsteadily the shock represented in those structures progresses, as it will accelerate for gradients steeper than $r^{-3}$ and decelerates for shallower slopes \citep{Sedo59}.
	The toroidal structures experience less growth, thus pre-existing toroidal R-T instabilities approach the remaining composition interfaces at a slower speed, making them less likely to penetrate the rear of the shock and the composition ``wall'' and spawn further instabilities.
	It has also been shown that slower plumes can lead to more interaction with the reverse shock, which further slows the bullets \citep{KiPlJa00,HaJaMu10,WoMuJa15}.
	
	\citet{WoMuJa15} improved upon the previous work of \citet{HaJaMu10} by running 3D shock breakout simulations with full $4\pi$ solid angle coverage of the star.
	Their simulations using four different progenitors allowed them to correlate the final morphologies to the different progenitor density structures.
	Although the metal-rich clumps were tied to the initial asymmetries of the explosion, they found that the shock and reverse shock dynamics determined by the density structure of the star were of prime importance in determining the final distribution of the ejecta.
	The methods described in \citet{WoMuJa15} were extended to generate light curves for potential progenitors of SN~1987A \citep{UtWoJa15,UtWoJa19, UtWoJa21}.
	
	Rayleigh-Taylor mixing in the context of CCSN shock breakout was further studied in \citet{MuGaHe18}, who ran 3D breakout simulations representing ultrastripped stars.
	They investigated the recent ideas proposed in \citet{Duff16} and \citet{PaScBa18} who theorized that R-T mixing could potentially be analyzed with a mixing-length treatment (MLT).
	It was found that a MLT does provide insight into the previously mentioned buoyancy versus drag dynamic.
	However, the simulations of \citet{MuGaHe18} suggest that MLT is insufficient to fully model R-T mixing in this problem.
	
	Finally, \citet{StJaKr20} also have performed full-sphere 3D shock breakout simulations with the aim of studying low-mass progenitors.
	This extensive study covered R-T mixing, morphology differences, ejecta composition, and remnant properties for all the evolutionary phases of the explosion.
	The full suite of 1D, 2D, and 3D model comparisons provide more evidence of the importance of the density structure of the progenitor star, as each model exhibited drastically different shock dynamics during the explosion.
	We use one of the same progenitors as in that study and compare our results below.
	
	While there has recently been a general trend in the community toward the development of self-consistent neutrino-driven explosions, only a select few groups follow the explosion all the way to shock breakout.
	The models presented here were run initially with the CCSN simulation code \chimera\ \citep{BrBlHi20}, which includes the relevant physics thought to be required to model the neutrino-driven CCSNe mechanism \citep{BrDiMe06,LeMeMe12a,LeMeMe12b}. 
	Accurate neutrino transport models that have successful explosions should lead to more reliable modeling of the ejecta in the explosion process. 
	
	To truly meet our goal of understanding the observable impacts of the central engine and R-T mixing on CCSN ejecta, simulations of the supernova explosion must be carried beyond the neutrino-driven phase where the central engine operates and the nucleosynthesis occurs.
	Until now, this is where \chimera\ models have ceased.
	Here, we take \chimera\ models as initial conditions to new simulations that follow the progression of the explosion through the entire star. 
	Utilizing self-consistent \chimera\ models, rather than parameterized models, provides the most faithful starting point currently available from a physics perspective.
	This is especially the case from a nucleosynthetic point of view.
	The previous shock breakout studies discussed above have only tracked, at most, a 13-species $\alpha$-network with two additional species to track beta decay and a composite tracer abundance for the rest of the iron peak species, while \chimera\ gives us the ability to track 160 nuclear species from $^{1}$H to $^{64}$Ge.
	
	We present a set of 3D simulations for a 9.6~\msun\ and a 10~\msun\ progenitor, both of which have already been exploded for the initial seconds in \chimera.
	As well as being nearly double both the radial and angular resolutions compared to \citet{WoMuJa15} and \citet{StJaKr20}, we present the first shock breakout simulations that evolve a large nuclear network (160 species).
	For one of the progenitors, however, only an axisymmetric model was available from the output of \chimera.
	Although it has been shown (see above) that the difference in using a 2D model versus a 3D model is significant due to the nature of the explosion mechanism, we have explored what utility a finished 2D model could provide in absence of a completed 3D model.
	This helps to ascertain the extent to which an axisymmetric model can be used in 3D to analyze observables at shock breakout.
	
	In a set of our axisymmetric simulations, we tilt the \chimera\ model on its axis 90\degr\ through a simple coordinate transform during mapping.
	The non-tilted axisymmetric models (2D \chimera\ models mapped directly to 3D in FLASH) mirror all quantities in $\theta$ throughout $\phi$.
	After applying the tilt transformation, $\phi$ velocities are introduced into the previously 100\% $\theta$ velocity system, which generates numerical perturbations, but conserves all grid quantities such as momentum and density.
	Through this, we explore how an axisymmetric model is able to deviate from its initial toroidal structure, thus behaving more like a true 3D explosion.
	
	Consequently, we have performed simulations in the following ways, with their respective naming conventions:
	\begin{enumerate}
		\item 2D \chimera\ model run in 2D within FLASH (only briefly discussed for comparison purposes).
		Referred to as D9.6--\twod, D10--\twod.
		\item 2D \chimera\ model launched with axisymmetry in 3D within FLASH.
		Referred to as D9.6--\spun, D10--\spun.
		\item 2D \chimera\ model launched with axisymmetry in 3D within FLASH, but tilted 90\degr\ counter-clockwise about the y-axis.
		Referred to as D9.6--\tilt,  D10--\tilt.
		\item 3D \chimera\ model, where available, run in 3D within FLASH.
		Referred to as D9.6--\threed.
	\end{enumerate}
	In Section~\ref{comp}, we describe the computational setup, input physics, as well as details about the progenitors.
	The progression of the explosion is detailed for the 9.6~\msun\ progenitor in Section~\ref{d96} and the 10~\msun\ progenitor in Section~\ref{d10}.
	Finally, we summarize our work and discuss key takeaways in Section~\ref{conc}.
	
	\section{Computational Setup} \label{comp}
	Our simulations were performed using the FLASH code \citep{FrOlRi00,DuAnGa09} developed by the Flash Center at the University of Chicago.
	FLASH has been used extensively to model Rayleigh-Taylor and associated instabilities, in both astrophysical and laboratory settings \citep{DiYoDi04,FiKaLa08,CoWhMi09,OnNaFe20}.
	
	The hydrodynamics are evolved using the explicit, directionally-split piecewise-parabolic method (PPM) to solve the compressible Euler equations.
	Although it is less sophisticated than some choices of hydrodynamics methods available in FLASH, the directionally-split PPM algorithm implements consistent multi-fluid advection \citep{PlMu99} that better maintains compositional gradients key to examining the distribution of isotopes in the ejecta.
	
	Self-gravity was included via FLASH's improved multipole solver that solves the Poisson equation through a multipole expansion \citep{CoGrFl13}.
	Although the 3D spherical multipole solver was not originally compatible with 3D spherical geometry in FLASH, a modified version of the solver was created for this work.
	
	\subsection{Grid Setup} \label{grid}
	Both two dimensional and three dimensional simulations were run in spherical geometry.
	The spherical geometry is natural for self-gravitating objects and allowed us to easily ``remove'' the region of space containing the proto-neutron star (PNS).
	Following the studies of \citet{WoMuJa15} and \citet{StJaKr20}, whose similar 3D simulations used parameterized models from the \hotb\ code \citep{ScKiJa06,ScJaFo08}, we use an inner radial boundary of 500~km to excise the PNS.
	Because of the high sound speed and fine zoning in that region, the excision helps alleviate the CFL time step constraint.
	A point mass was placed at the origin to replace the mass of the excised PNS.
	
	These simulations are intended to accurately capture the explosion throughout the entire star, approximately $10^{8}$~km in radius.
	An efficient way to accomplish this in spherical coordinates is to use a logarithmically-spaced radial grid, as described in \citet{Fern12} and shown in \citet{WoMuJa15}.
	This type of grid provides the ability to more easily maintain ``square'' zones with constant $\Delta \theta \approx \Delta r/r$, and can more accurately track the near power law density structure of stars.
	Though adaptive mesh refinement (AMR) provides an excellent way to resolve specific regions of the explosion, while efficiently ignoring others, FLASH's AMR is incompatible with log spacing.
	Consequently, we have implemented a log-spaced version of FLASH's uniform grid using logarithmically spaced blocks and uniformly-spaced cells within each block along the radial dimension.
	
	As outlined in \citet{Fern12}, we similarly define the domain between $r_{\rm min}$ and $r_{\rm max}$ such that consecutive block sizes have a ratio $\Delta r_{i+1}/\Delta r_{i} = \zeta > 1$, where $i$ is the block number, which increases with increasing radius.
	Logarithmic block spacing is achieved by setting
	\begin{equation}
	\zeta = (r_{\rm max} /r_{\rm min} )^{1/N_{r}} \, ,
	\end{equation}
	where $N_{r}$ is the number of radial blocks.
	The grid is then created over $0 \leq q \leq N_{r}$ by defining the inner edge of each block as:
	\begin{equation}
	r_{q} = \zeta ^{q} \; r_{\rm min} \, ,
	\end{equation}
	where $r_{0} = r_{\rm min}$ and $r_{N_{r}} = r_{\rm max}$ .
	Each logarithmically spaced block contains 16 uniformly spaced cells in the radial direction.
	
	The inner and outer radial grid boundaries are diode and outflow, respectively, the polar grid boundaries are reflecting, and the azimuthal boundaries are periodic.
	The diode boundary condition is similar to outflow, but only allows matter to flow out of the domain, as opposed to letting matter freely enter the domain as well.
	The inner boundary is fixed until the first R-T instabilities begin to develop ($\sim$2--3 seconds), then is shifted to larger radii, following the progress of the shock.
	This is accomplished by removing the innermost radial block whenever the inner boundary becomes smaller than 1\% of the minimum shock radius.
	This removes the region where the PNS, absent from our model, may have influence in the form of a PNS wind, and makes the simulation computationally cheaper, progressively reducing the number of radial zones, which, in turn, relaxes the CFL time step constraint.
	The mass loss caused by moving the inner boundary is small, but not negligible -- it is consistently $\sim$10$^{-5}$~\msun\ in all of our 3D simulations.
	As we will discuss in Section~\ref{d96tilt}, this accounts for only $\sim$1.5\% of the total mass lost, while the rest is due to fallback (matter passing through the inner boundary).
	
	In addition, for all simulations, we define angle-averaged grid quantities as:
	\begin{equation}
	\langle X(r) \rangle = \frac{\int X(r) \, d\Omega}{\int d\Omega} \, ,	
	\end{equation}
	where $d\Omega$ is the differential of the solid angle.
	
	\subsection{Grid Numerics} \label{gridnum}
	For each 3D model, the grid initially consists of $2304\times192\times384$ total cells in $r,\theta,\phi$, respectively.
	The radial section of the grid extends logarithmically from 500~km $\leq r \leq R_{\star}$, where $R_{\star}$ is the stellar radius of the progenitor (see Table~\ref{tab:prog}).
	As with \citet{HaJaMu10}, cones were excised along the poles to help further relax the CFL condition in 3D -- in this case having a half-opening angle of 5\degr. 
	The grid therefore covers $0.0278 \pi \leq \theta \leq 0.972 \pi$ at $\delta \theta = 0.885$\degr\ and azimuthal angles $0 \leq \phi \leq 2 \pi$ at $\delta \phi = 0.938$\degr.
	In two dimensions, the grid covers the same radial extent and polar angles $0 \leq \theta \leq \pi$ with $2304\times204$ cells, respectively.
	
	This leads to a radial resolution of nearly constant $\Delta r/r$ of $5.7 \times 10^{-3}$ and $6.1 \times 10^{-3}$ for models D9.6 and D10, respectively.
	This can be compared to $6.9 \times 10^{-3}$ of \citet{MuGaHe18}, $8.9 \times 10^{-3}$ of \citet{StJaKr20}, and $1 \times 10^{-2}$ for \citet{HaJaMu10} and \citet{WoMuJa15}.
	All of our 3D runs have angular resolutions $<1\degr$, compared to $1\degr$ for \citet{HaJaMu10}, $1.6\degr$ for \citet{MuGaHe18}, and $2\degr$ for both \citet{WoMuJa15} and \citet{StJaKr20}.
	
	\begin{deluxetable}{lcccl}
		\tablecaption{Progenitor Structure \label{tab:prog}}
		\tablecolumns{5}
		\tablewidth{0pt}
		\tablehead{
			\colhead{Model} & \colhead{(C+O)/He} & \colhead{He/H} & \colhead{$R_{\star}$} & \colhead{\tmap \tablenotemark{a}} \vspace{-.2cm} \\
			\colhead{ } & \colhead{[km]} & \colhead{[km]} & \colhead{[km]} & \colhead{[s]}
		}
		\startdata
		D9.6 & \ten{6.95}{3} & \ten{1.40}{7} & \ten{1.50}{8} & 0.650 (0.467) \\
		\hline
		D10 & \ten{2.02}{4} & \ten{4.32}{6} & \ten{3.57}{8} & 1.763 (...)  \\
		\enddata
		\tablenotetext{a}{Mapping time of the 3D \chimera\ model given in parentheses.}
		\tablecomments{Radii of the composition interfaces are defined as the positions at the edge of the stellar layers where the dominant mass fraction of the layer drops below half its maximum value within the layer. $R_{\star}$ is the stellar radius of the progenitor, and \tmap\ indicates the post-bounce time when the \chimera\ conditions are mapped into FLASH.}
	\end{deluxetable}
	\newpage
	
	\subsection{Equation of State} \label{eos}
	The removal of the PNS from the grid allows us to neglect the high densities and temperatures present there and use FLASH's implementation of the Helmholtz equation of state \citep[Helmholtz EoS;][]{TiSw00}, which displays perfect thermodynamic consistency and includes contributions to internal energy from ions, electrons, positrons, and radiation.
	Because the Helmholtz EoS assumes full ionization, we halt each simulation when the shock front reaches the region of the progenitor where this criterion is no longer true ($T \lesssim 10000$ K), which happens only a few zones before shock breakout for our models.
	
	\subsection{Nuclear Network} \label{net}
	FLASH allows us to track large numbers of species and utilize a multispecies network for nuclear burning, in this case the FLASH implementation of \xnet\ \citep{HiTh99a}.
	Nuclear burning does not occur during these extended FLASH runs unless the \chimera\ runs serving as our initial conditions were stopped before nuclear burning was complete, which is the case with the D9.6--\threed\ model, where \xnet\ was on for the first 279 milliseconds.
	Regardless, the inclusion of \xnet\ gives us the unprecedented ability to track the composition of 160 nuclear species throughout the evolution of the explosion, which leads to a more accurate analysis of the ejecta seen at shock breakout.
	The species list of the {sn160} network from \xnet\ is given in Table~\ref{tab:network}.
	Of particular note are \isotope{Ge}{64} and \isotope{Zn}{66}, where proton-rich and neutron-rich flows that would progress to higher atomic number in nature stagnate in this network.
	
	\subsection{Initial Conditions and Progenitor Models} \label{init}
	Each of our FLASH simulations are initialized from the final step of a 2D or 3D \chimera\ neutrino radiation hydrodynamics simulation of the supernova mechanism that has reached an asymptotic explosion energy.
	The \chimera\ simulations are initialized from two 1D progenitors calculated with the \kepler\ stellar evolution code \citep{WeZiWo78} up to the moment of Fe-core collapse.
	These simulations are part of \chimera's `D-series' (and so prefixed) and are substantially similar in input physics to prior \chimera\ simulations except for the inclusion of the larger 160-species nuclear network to better handle the formation of neutron-rich ejecta.
	The explosion dynamics of the input \chimera\ simulations will be described in forthcoming publications.
	As the \chimera\ simulations do not include parts of the outer core or the envelope of the progenitor, the missing portions of the \kepler\ progenitor are reattached outside the limit of each \chimera\ simulation when mapping to FLASH.
	The \chimera\ simulations are ended when the computational intensity required to simulate the neutrino mechanism is no longer needed.
	
	\begin{deluxetable}{cccccc}
		\tablecaption{Species List \label{tab:network}}
		\tablewidth{0pt}
		\tablehead{ \multicolumn{6}{c}{Species in sn160 network} }
		\startdata
		\isotope{n}{} & \isotope{H}{1\textrm{--}2} & \isotope{He}{3\textrm{--}4} & \isotope{Li}{6\textrm{--}7} & \isotope{Be}{7,9} & \isotope{B}{8,10,11}  \\
		\isotope{C}{12\textrm{--}14} &\isotope{N}{13\textrm{--}15} & \isotope{O}{14\textrm{--}18} & \isotope{F}{17\textrm{--}19} & \isotope{Ne}{18\textrm{--}22}& \isotope{Na}{21\textrm{--}23}\\
		\isotope{Mg}{23\textrm{--}26} & \isotope{Al}{25\textrm{--}27} &  \isotope{Si}{28\textrm{--}32}&  \isotope{P}{29\textrm{--}33} & \isotope{S}{32\textrm{--}36} &\isotope{Cl}{33\textrm{--}37} \\
		\isotope{Ar}{36\textrm{--}40} & \isotope{K}{37\textrm{--}41} & \isotope{Ca}{40\textrm{--}48} & \isotope{Sc}{43\textrm{--}49} & \isotope{Ti}{44\textrm{--}51} & \isotope{V}{46\textrm{--}52} \\
		\isotope{Cr}{48\textrm{--}54} & \isotope{Mn}{50\textrm{--}55} & \isotope{Fe}{52\textrm{--}58} & \isotope{Co}{53\textrm{--}59} & \isotope{Ni}{56\textrm{--}64} & \isotope{Cu}{57\textrm{--}65} \\ 
		\isotope{Zn}{59\textrm{--}66} & \isotope{Ga}{62\textrm{--}64} & \isotope{Ge}{63\textrm{--}64} \\
		\enddata
	\end{deluxetable}
	
	\begin{figure*}
		\includegraphics[width=2.1\columnwidth,clip]{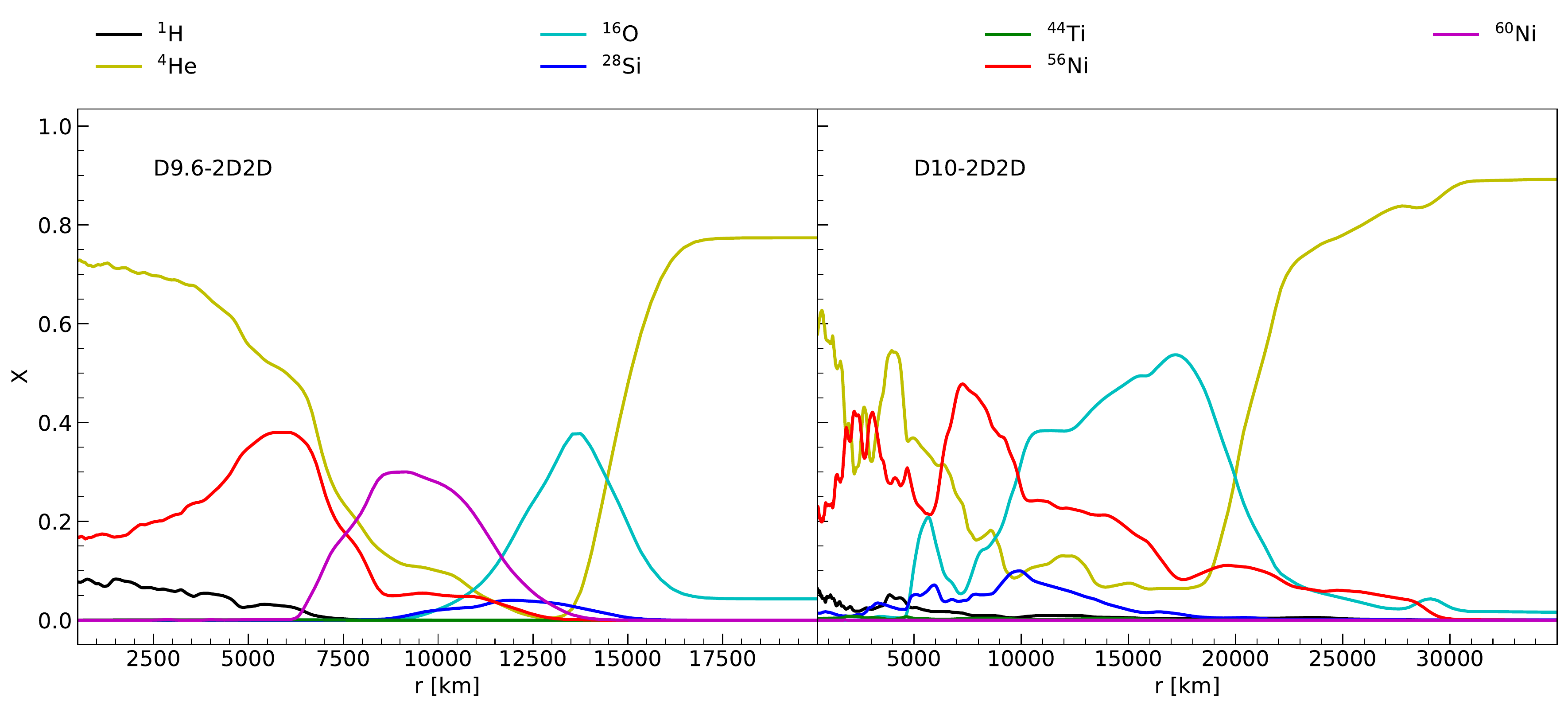}
		\caption{\label{fig:compo}
			Angle-averaged mass-fractions of inner ejecta for D9.6--\twod\ (left) and D10--\twod\ (right) at \tmap.
		}
	\end{figure*}
	
	\begin{figure*}
		\includegraphics[width=2.1\columnwidth,clip]{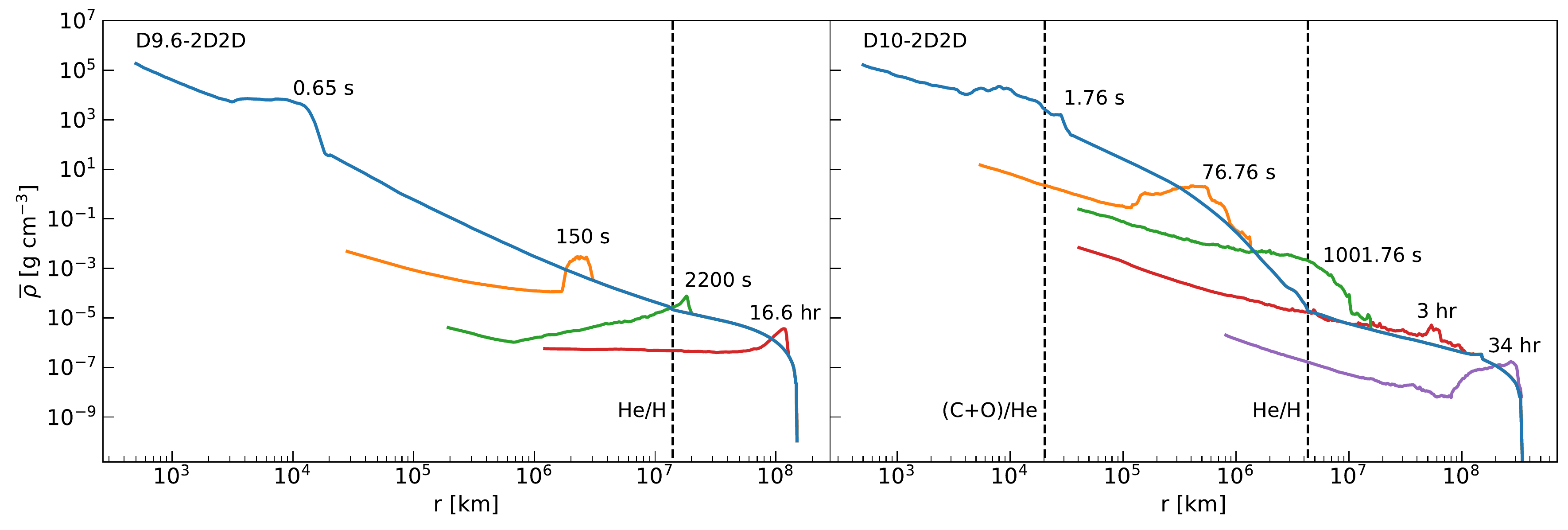}
		\caption{\label{fig:dens_both}
			Left: Evolution of the angle-averaged density profile for the D9.6--\twod\ model. The vertical dashed line indicates the position of the He/H interface. Right: Evolution of the angle-averaged density profile for the D10--\twod\ model. The vertical dashed lines indicate the positions of the (C+O)/He and He/H interfaces.
		}
	\end{figure*}
	
	The first progenitor is a 9.6~\msun\ zero-metallicity star, provided by A. Heger (private communication).
	That choice is motivated by 3D simulations of this same progenitor by other groups \citep{MeJaMa15,MuTaHe19,StJaKr20} and is used to explore progenitors relating to low-mass cores, and to demonstrate the 160 species network.
	The diagnostic explosion energies at the end of the \chimera\ runs are \ten{1.91}{50}~ergs for \chimera\ model `D9.6-sn160-2D' and \ten{1.68}{50}~ergs for \chimera\ model `D9.6-sn160-3D', where the difference is due, in part, to the 2D run having been evolved nearly 200~ms further than the 3D model (E.J. Lentz et al., in prep.).
	We refer to these models collectively as D9.6 in this work.
	The explosion energy in 3D is $\sim$95\% higher than the energy reported for the same progenitor in \citet{StJaKr20} at the time of their mapping from \vertex\ to \hotb.
	Like low-mass oxygen-neon supernovae, the shock did not stall after bounce and the initial ejecta includes neutron-rich material drawn from the vicinity of the PNS that is not exposed to neutrino radiation in protracted pre-explosion convective heating.
	As a result, the outer ejecta of this explosion is enhanced with neutron-rich species like \isotope{Ca}{48}, \isotope{Ni}{60}, and \isotope{Zn}{66} that are not seen in typical iron-core CCSNe and less of isotopes like \isotope{Ni}{56} and \isotope{Ti}{44} that are more common in supernovae that take longer to explode.
	This neutron-rich ejecta is noticeable at \tmap\ in Figure~\ref{fig:compo} (left), and can be seen in the \isotope{Ni}{60} peak that rivals typical CCSN ejecta like \isotope{Ni}{56}.
	
	At the time it is mapped into FLASH, the D9.6 model has a relatively featureless density profile which gradually decreases ahead of the shock front until the edge of the star (Figure~\ref{fig:dens_both}, left).
	The mean shock position at \tmap\ resides in the He shell, which extends from \ten{6.95}{3}~km to \ten{1.40}{7}~km, and accounts for 0.33~\msun\ of the total mass.
	An extensive H-envelope spans from the edge of the He shell to the edge of the star at \ten{1.50}{8}~km, and accounts for 7.85~\msun\ of the total mass.
	The compactness parameters, as described in \citet{OcOt11}, are $\xi_{2.5} = 7.65\times10^{-5}$ and $\xi_{1.5} = 2.34\times10^{-4}$, which are smaller overall compactness than our second progenitor.
	
	The second progenitor, a 10~\msun\ solar-metallicity star, was presented in \citet{SuErWo16} as a part of their study of 200 pre-supernova models.
	The 2D \chimera\ model (`D10-sn160-SEWBJ16', J.A. Harris et al., in prep.) has a diagnostic explosion energy of \ten{3.075}{50} ergs, which is almost double the energy of the D9.6 3D at its \tmap and $\sim$50\% higher than the D9.6 2D energy.
	
	The D10 \chimera\ model is a traditional CCSN model with the shock stalling shortly after bounce and significant accretion onto the PNS occurring.
	This explains the lack of neutron-rich material when comparing composition profiles of the two models in Figure~\ref{fig:compo} --- note the lack of a \isotope{Ni}{60} peak.
	Combined with a significantly higher presence of \isotope{C}{12} and \isotope{O}{16}, this leads to a different profile in the ejecta lying behind the shock. 
	More fluctuations are also noticeable within the composition profile due to the more prolate shock front in the D10 compared to the relatively spherical shock front of the D9.6, leading to a less uniform angular distribution of ejecta.
	
	In further contrast to the D9.6, this progenitor has a rather erratic density profile, especially noticeable in $\rho r^{3}$, with a dramatic change in density gradient at the He/H interface (Figure~\ref{fig:dens_both}, right).
	The mean shock position at \tmap\ resides in the former He-burning shell, which extends from \ten{2.02}{4}~km to \ten{3.20}{5}~km.
	This shell is a key feature in the $\rho r^{3}$ profile, for it is the source of a dramatic acceleration that the shock experiences when transitioning to the inert He layer residing above.
	The He-burning shell contributes 0.44~\msun\, while the remaining He layer, which ends at \ten{4.32}{6}~km, provides a comparable 0.43~\msun\ for a total mass of 0.87~\msun\ of the entire He shell.
	The similar masses for each section of the He shell spread across widely different spatial extents explains the change in density gradient at the transition point.
	A hydrogen envelope spans from the edge of the He shell to the edge of the star at \ten{3.57}{8}~km, and accounts for 7.2~\msun\ of the total mass.
	An additional density feature can be seen near the edge of the hydrogen shell located at \ten{1.49}{8}~km.
	The compactness parameters are $\xi_{2.5} = 2.04\times10^{-4}$ and $\xi_{1.5} = 4.32\times10^{-1}$.
	The large difference between $\xi_{2.5}$ and $\xi_{1.5}$ is the result of the (C+O)/He interface lying at 1.61 \msun.
	Details of key interfaces for both progenitors, as well as mapping times from \chimera\ to FLASH, are given in Table~\ref{tab:prog}.
	
	\section{D9.6 - Results} \label{d96intro}
	In this section, we discuss the general progression of the shock in the D9.6--\threed\ model (Section~\ref{d96}), with slight deviations to the story, specific analysis, and comparisons to D9.6--\spun\ and D9.6--\tilt\ residing in Sections~\ref{d96spin} and \ref{d96tilt}.
	Lastly, we compare our models to previous studies of the same progenitor in Section~\ref{d96compare}.
	
	\begin{figure}
		\includegraphics[width=\columnwidth,clip]{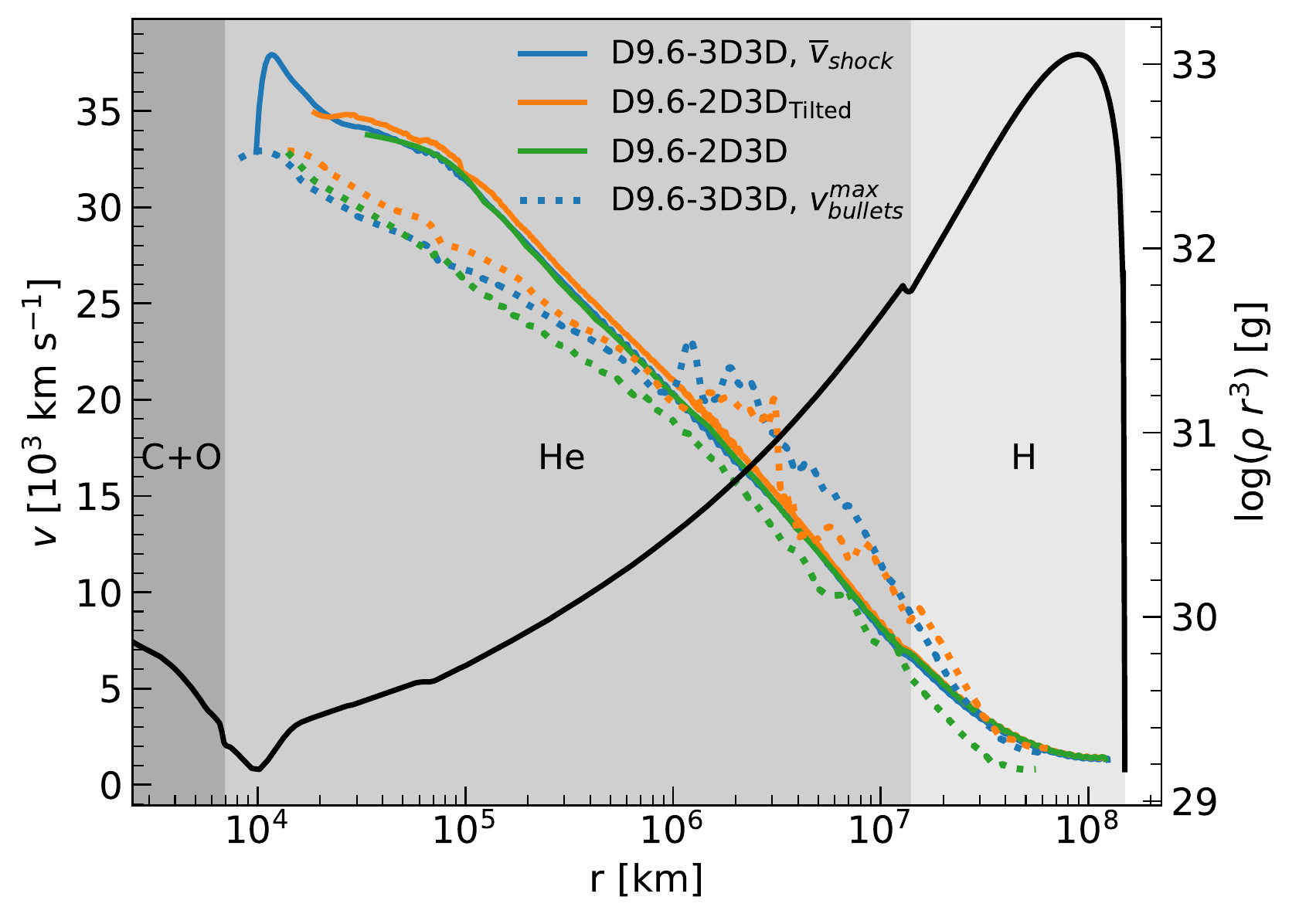}
		\caption{\label{fig:d96shockvel}
			Angle-averaged shock velocity (colored, solid lines) and maximum velocity of the $X_{\rm \isotope{Ni}{56}+IG} = 3\% $ bullet isosurface (colored, dashed lines) for the D9.6 models as functions of their respective angle-averaged shock or bullet radii. Density profile of the D9.6 progenitor prior to bounce (black, solid) displays the change of $\rho r^{3}$ and spans the right axis. Grey shaded sections highlight the regions of the (C+O), He, H shells up to the defined interfaces in Table~\ref{tab:prog}.
		}
	\end{figure}
	
	\subsection{D9.6-3D3D} \label{d96}
	Model D9.6--\threed\ was mapped into FLASH at a time \tmap\ $=$ 466.6~ms after the bounce that marks the formation of the PNS, having been simulated to that point with \chimera.
	At this point in the explosion, the mean shock radius is at $\sim$\ten{1.0}{4}~km, just across the (C+O)/He interface.
	As noted by \citet{StJaKr20}, this progenitor is in the process of a 2nd dredge-up of the He shell which has created a section at the base of the shell that contains minimal hydrogen (in contrast to the rest of the He shell).
	The shock encountering changes to $\rho r^{3}$ in this region explains the slight deviation in the trend of the shock velocity at $\sim$\ten{1.7}{4}~km and $\sim$\ten{6.0}{4}~km (see Figure~\ref{fig:d96shockvel}).
	
	\begin{figure*}
		\includegraphics[width=509.76pt]{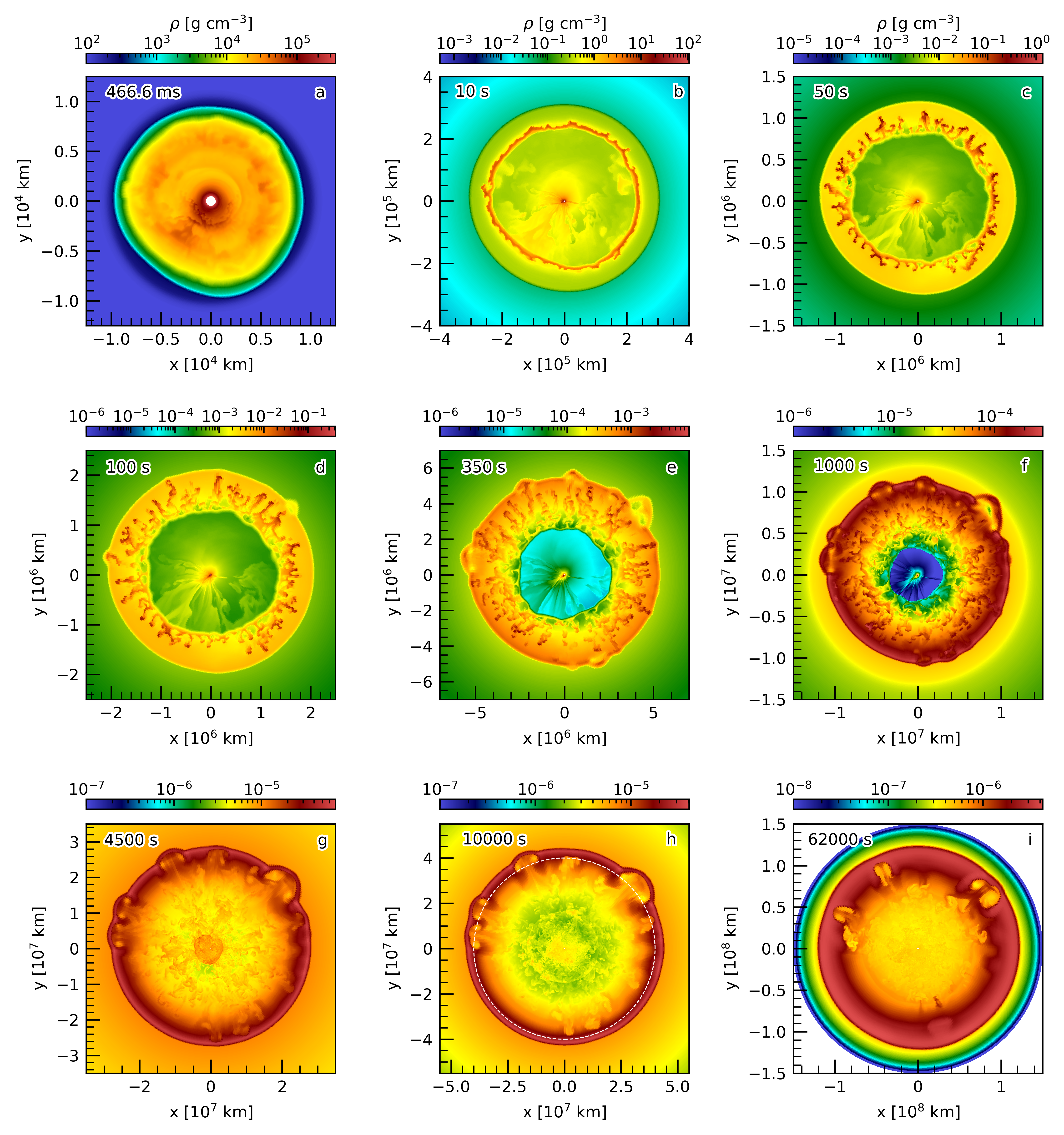}
		\caption{\label{fig:d96rhoslices}
			Slices of density in the D9.6--\threed\ model at displayed times. Note the dramatic changes between the panels in the axis limits and the color bar range as the ejecta expands. Green to yellow color discontinuity ahead of the shock in panel (f) represents the position of the He/H interface. In panel (g), the secondary blast wave that resulted from the rebound of the first reverse shock (the collapsing blue region behind the shock in prior panels) is visible at $\sim$\ten{5.0}{6}~km. White dashed circle in panel (h) marks the \ten{4.0}{7}~km radius used to slice the plumes in Figure~\ref{fig:d96plume_frac}.
		}
	\end{figure*}
	
	\begin{figure*}
		\includegraphics[width=509.76pt]{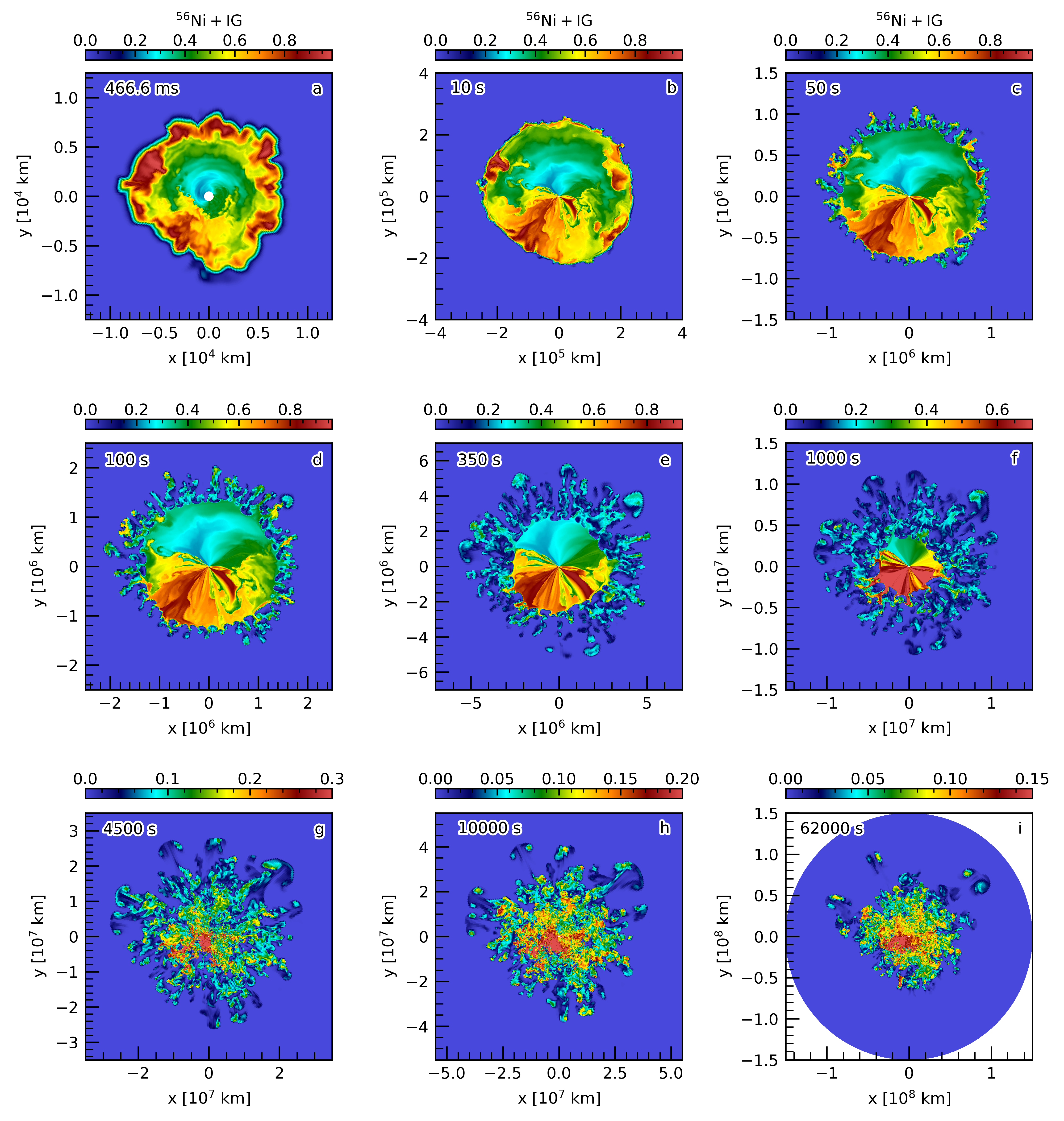}
		\caption{\label{fig:d96bulletslices}
			Slices of $X_{\rm \isotope{Ni}{56}+IG}$ in the D9.6--\threed\ model at displayed times (same times and axes as in Figure~\ref{fig:d96rhoslices}).  Note change in color bars in later panels as the heavy elements become diluted.
		}
	\end{figure*}

	\begin{figure*}
	\includegraphics[width=509.76pt]{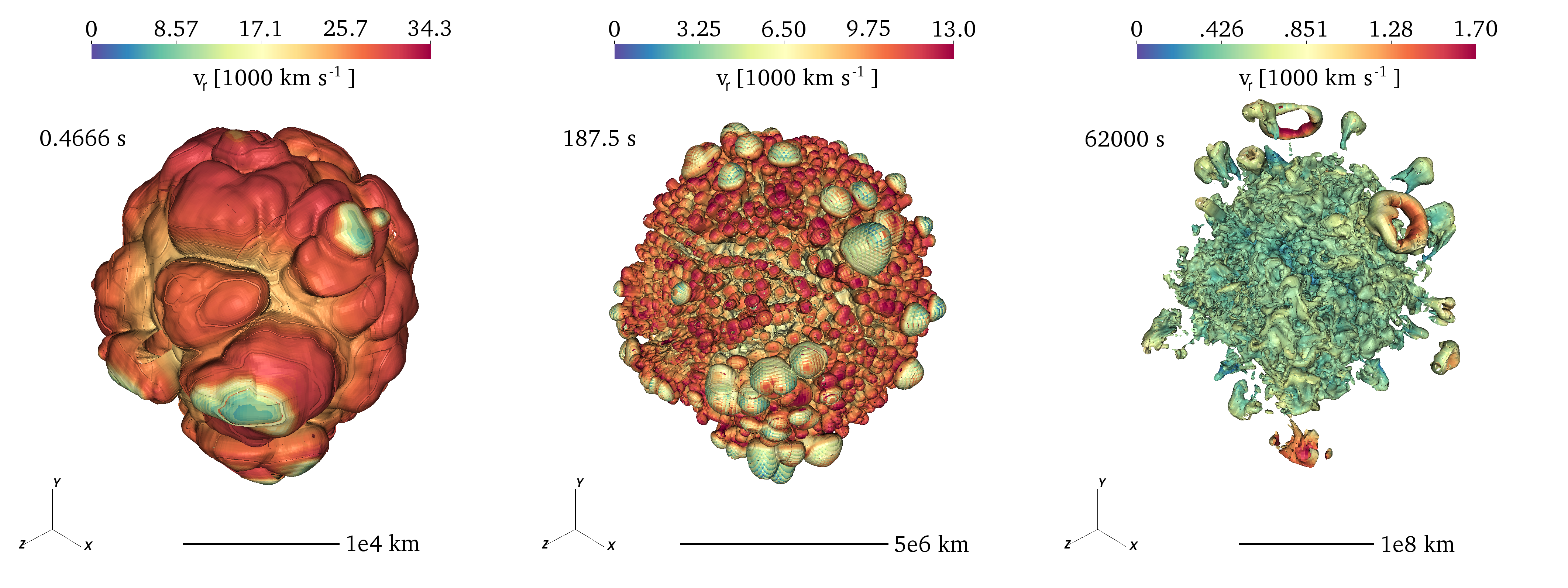}
	\caption{\label{fig:d96velr}
		Time snapshots of the $X_{\rm \isotope{Ni}{56}+IG} = 3\%$ isosurface (color-coded by radial velocity) in the D9.6--\threed\ simulation. Initial asymmetries at \tmap\ (left) evolve into mushroom features in the He shell (middle) that seed further R-T bullets seen at shock breakout (right).
	}
	\end{figure*}

	At the time of mapping from \chimera\ to FLASH, the metal-rich shell lying behind the shock is mainly composed of \isotope{C}{12} and \isotope{O}{16}.
	It is this shell, which is quasi-spherical, that begins to deform and starts to develop the initial R-T instabilities.
	This shell, located at the green to yellow transition in Figure~\ref{fig:d96rhoslices}(a) at $\sim$\ten{8.5}{3}~km, is also the location of the reverse shock created upon crossing the (C+O)/He interface.
	The departure from sphericity is imprinted on the reverse shock at its creation.
	Although the location of the mass shell is the position of the reverse shock in this scenario, this is not always the case as we will see with the He/H interface discussed below.
	Because the main shock has only just crossed the (C+O)/He interface, there are still portions of it that are still traveling down the density cliff, thus overall the shock is still accelerating at this point --- represented by the velocity spike shown in Figure~\ref{fig:d96shockvel}.
	There is also a wind-termination shock \citep[also in][]{StJaKr20} that resides close to the inner boundary and will eventually collapse inwards due to the absence of the PNS and its wind from the simulation.
	From this point forward, the explosion propagates through the He core, and the deformed metal-rich shell starts to mix with that material. 
	
	By $\sim$2~s, the shock has crossed fully into the He layer with the initial R-T plumes appearing as ripples in the fragmenting metal-rich shell.
	The instabilities begin to develop their typical mushroom state at 10~s and are still mainly composed of the species from the metal-rich shell (see ripples at $\sim$\ten{2.3}{5}~km in Figures~\ref{fig:d96rhoslices}(b), \ref{fig:d96bulletslices}(b)). 
	Beginning at approximately 30~s, the inner regions of the ejecta (the ``hot bubble'') are injected into the rear of the instabilities, including the key isotopes of nickel like \isotope{Ni}{56} and \isotope{Ni}{60}.
	To track the bullets, we have combined the mass fractions of \isotope{Ni}{56} and neutron-rich iron group nuclei ($X_{\rm \isotope{Ni}{56}+IG}$) and have taken a 3\% isosurface of the result, which enables a direct comparison to the tracking of $X_{\rm \isotope{Ni}{56}+Tr}$ bullets in \citet{StJaKr20}.
	To approximate the crude tracer nucleus of \citet{StJaKr20}, we define ``neutron-rich iron group'' as all species in our network falling in the range of \isotope{Cr}{49}--\isotope{Ni}{64}, excluding \isotope{Fe}{52} and \isotope{Ni}{56}.
	
	Large-scale features start to form at $\sim$60~s where the radial shock position is $\sim$\ten{1.0}{6}~km, which can be seen in Figure~\ref{fig:d96shockvel} as the $X_{\rm \isotope{Ni}{56}+IG}$ maximum velocity (dashed blue) curve crosses the shock's (solid blue) curve.
	Fluctuations in the velocity after the crossing point are due to plume interactions with the shock.
	As the fastest moving bullet penetrates the shock, that bullet slows, and the maximum velocity shifts to the next fastest bullet.
	This continues until all of the fast moving clumps eventually interact with the shock front, which then results in a steady decline of the maximum velocity of these R-T plumes.
	These features can explicitly be seen penetrating the shock in Figure~\ref{fig:d96rhoslices}(d).
	
	By 150~s, there is no semblance left of the metal-rich shell, as the inner ejecta from the hot bubble has completely engulfed it.
	The R-T fingers have grown significantly by this point and have reached the back of the shock (see large mushroom features in middle panel of Figure~\ref{fig:d96velr}).
	As the shock continues to progress through the He core, the R-T fingers progress with it, remaining near the rear of the shock (see elongated fingers penetrating the shock in Figures~\ref{fig:d96rhoslices}(e) and \ref{fig:d96bulletslices}(e)).
	Whether the R-T fingers penetrate the shock is key to the morphology of the remnant.
	The shock experiences a gradual deceleration in this region of the progenitor due to the  increasing $\rho r^{3}$ and the extent of the He shell.
	Additionally, the reverse shock created at the first density interface has continued to propagate inward in mass and starts to shred the inner regions (the blue region at $\sim$\ten{2.5}{6}~km in Figure~\ref{fig:d96rhoslices}(e)). 
	This reverse shock is not spherical as a consequence of the asphericity and timing of the main shock's interactions with the prior composition interface.
	
	\begin{figure*}
		\includegraphics[width=2.1\columnwidth,clip]{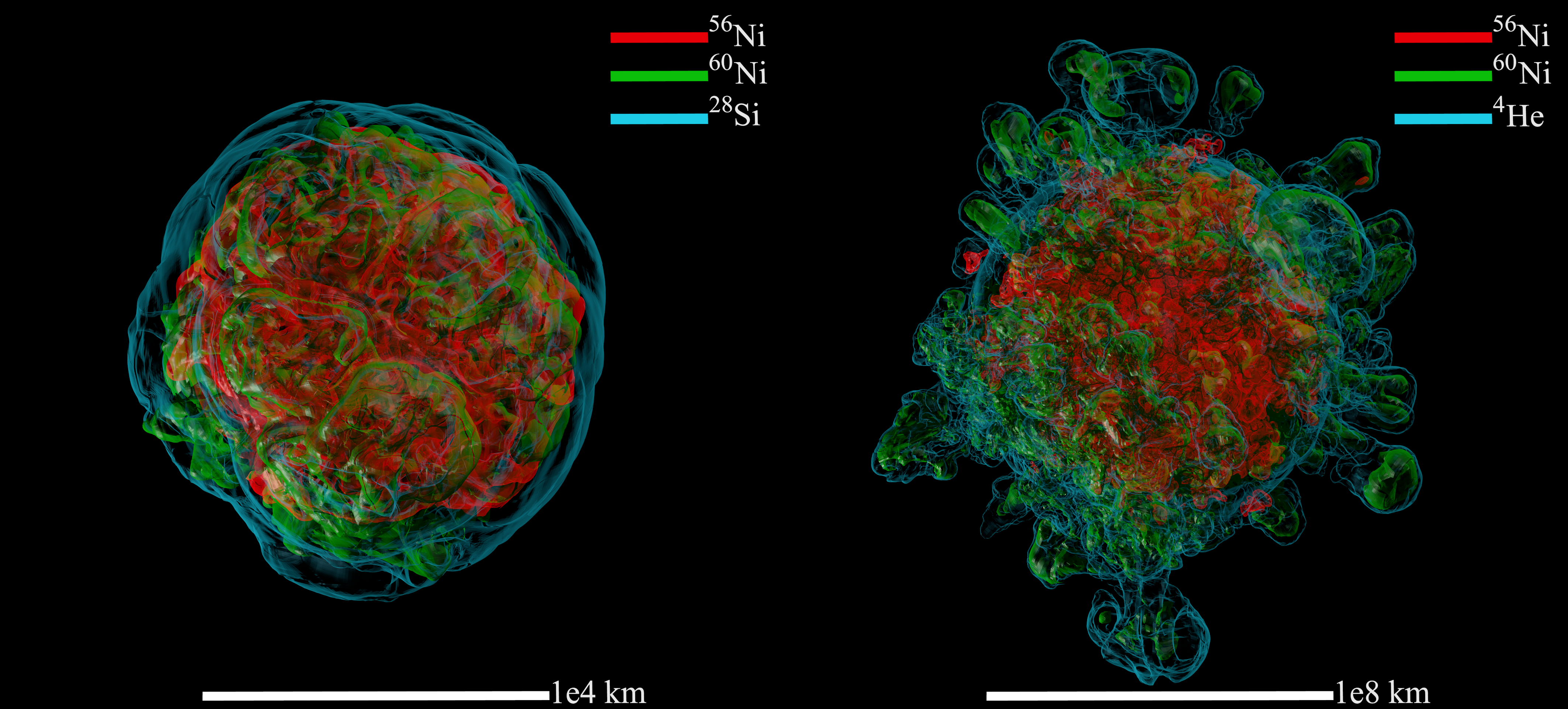}
		\caption{\label{fig:d96nini}
			Left: Isosurfaces at 5\% mass fraction of \isotope{Ni}{56} (red) and \isotope{Ni}{60} (green) reveal the early morphology of inner ejecta surrounded by a shell of \isotope{Si}{28} (cyan) displayed as a 1\% mass fraction isosurface for the D9.6--\threed\ model at \tmap\ (466.6~ms). Right: Isosurfaces at 1\% mass fraction of \isotope{Ni}{56} (red) and \isotope{Ni}{60} (green) highlight the inner anatomy of the \isotope{He}{4} (cyan) coated bullets displayed as a 40\% mass fraction isosurface for the D9.6--\threed\ model at shock breakout ($\sim$62000~s).
		}
	\end{figure*}
	
	At $\sim$1000~s (Figures~\ref{fig:d96rhoslices}(f) and \ref{fig:d96bulletslices}(f)), the shock crosses the He/H composition interface located at \ten{1.4}{7}~km and creates a weak pressure wave due to the minimal change in $\rho r^{3}$ (see Figure~\ref{fig:d96shockvel}) that propagates inward in mass and radius before eventually steepening into a second reverse shock.
	This delay ensures that the second reverse shock location is quickly decoupled from position of the mass shell at the He/H interface, whereas the (C+O)/He mass shell and first reverse shock positions coincided.
	Although slight, the deceleration gives the closest R-T plumes to the shock front the opportunity to interact with the rear of the shock.
	We only see this interaction happen in D9.6--\threed\ and D9.6--\tilt, which end up having higher overall velocities compared to the other simulations (discussed in Sections~\ref{d96spin} and \ref{d96tilt}).
	This interaction not only seeds new instabilities, but it further develops the most dominant R-T fingers into even larger mushroom-shaped plumes that are able to penetrate and re-shape the shock.
	These will be the fastest bullets at shock breakout, though the shock must still propagate through most of the envelope before they reach that point.
	
	By 2500~s, the inner regions of the explosion are completely shredded by the first reverse shock.
	The mixing effects can be seen when transitioning from Figure~\ref{fig:d96bulletslices}(f) to \ref{fig:d96bulletslices}(g) (note that the distribution of ejecta is much less uniform in the inner regions after the transition).
	Having been born quasi-spherical and propagated through inhomogeneous regions of the star leaves the reverse shock aspherical.
	As a result, although some portions of the collapsing reverse shock pass through the inner radial boundary (located at $r =$~\ten{1.8}{5}~km at this point in the simulation), most of the reverse shock bypasses the boundary altogether and collides with itself off center rather than at the PNS.
	This sets up the creation of a secondary forward-propagating blast wave which is reminiscent of the implosions discussed in supernova remnant theory \citep{TrMc99,CiMcBe88}.
	The blast wave can be seen at $\sim$\ten{5.0}{6}~km in Figure~\ref{fig:d96rhoslices}(g).
	This causes significantly more mixing, as the inner regions also bounce off the reflecting boundaries of the grid, and the R-T plumes grow to be quite abundant.
	
	What once were primarily metal-rich mushrooms are now heavily coated in helium, for the propagation through the He core has filled the gaps between the R-T fingers and has shaped them further.
	However, as noted earlier, the original inner regions of the ejecta still form the ``bulk'' of the inner anatomy of a single finger (see Figure~\ref{fig:d96nini}) due to the injection through the metal-rich shell.
	Most notably represented in the main anatomy of an instability are the Ni isotopes, as expected, with the most abundant isotope occupying the bullets being $^{60}$Ni, from the early, neutron-rich portion of the hot bubble.
	
	As the shock continues to expand, the pressure wave created at the He/H interface reaches the center of the grid at $\sim$20000~s, after steepening into a reverse shock at $\sim$16000~s.
	As the first reverse shock collided with itself, so does the second, but it does not rebound as hard as the former, because this second reverse shock has been weak since its launch as a result of the slight deceleration of the main shock noted above.
	Nevertheless, reaching the center still creates another forward-propagating blast wave which further influences the inner ejecta.
	
	\begin{figure}
		\includegraphics[width=\columnwidth,clip]{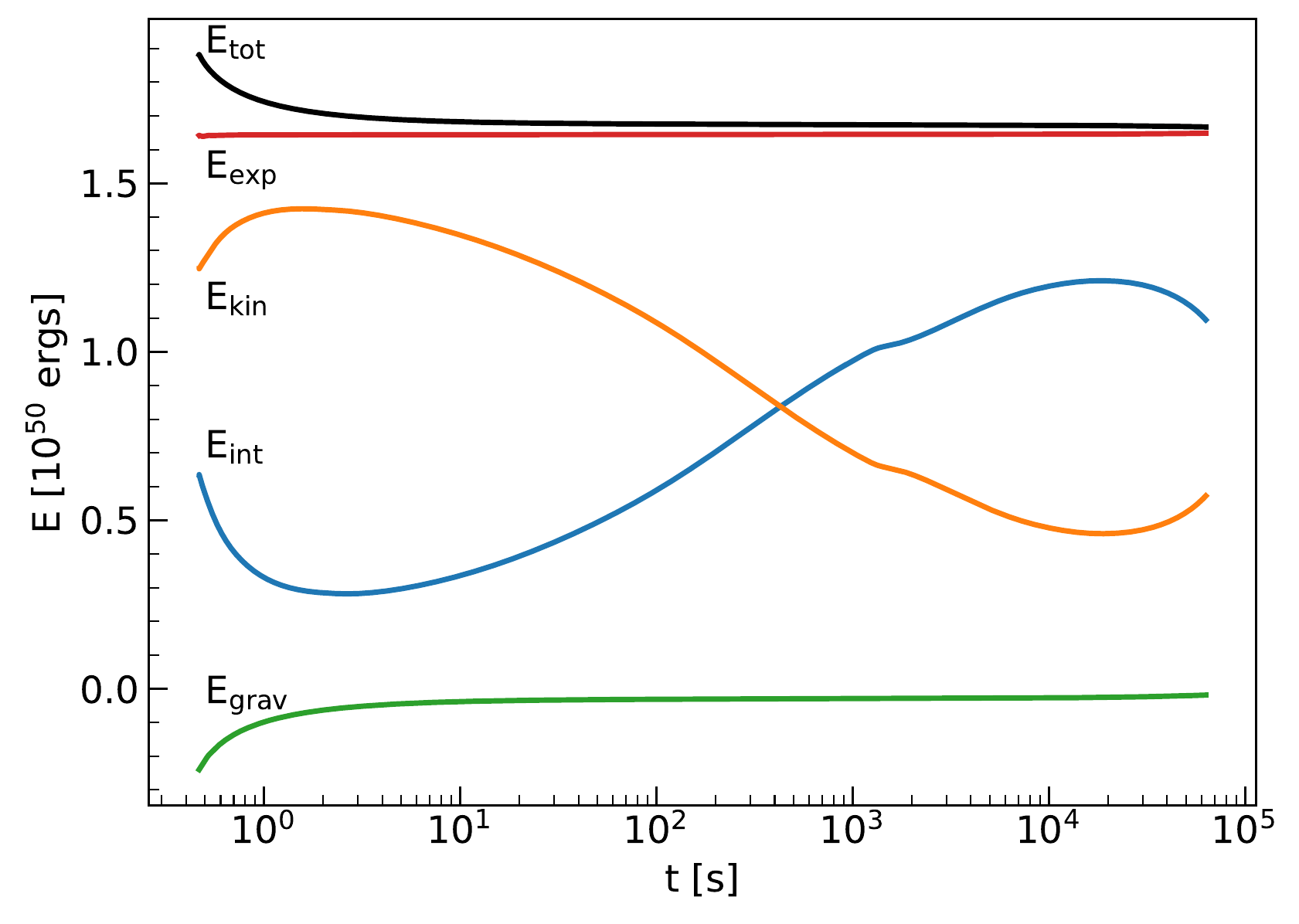}
		\caption{\label{fig:d96energy}
			Evolution of total energy (black), explosion energy (red), kinetic energy (orange), internal energy (blue), and gravitational binding energy (green) in the D9.6--\threed\ model. 
		}
	\end{figure}
	
	\begin{figure*}[p]
		\centering
		\includegraphics[width=1.89\columnwidth]{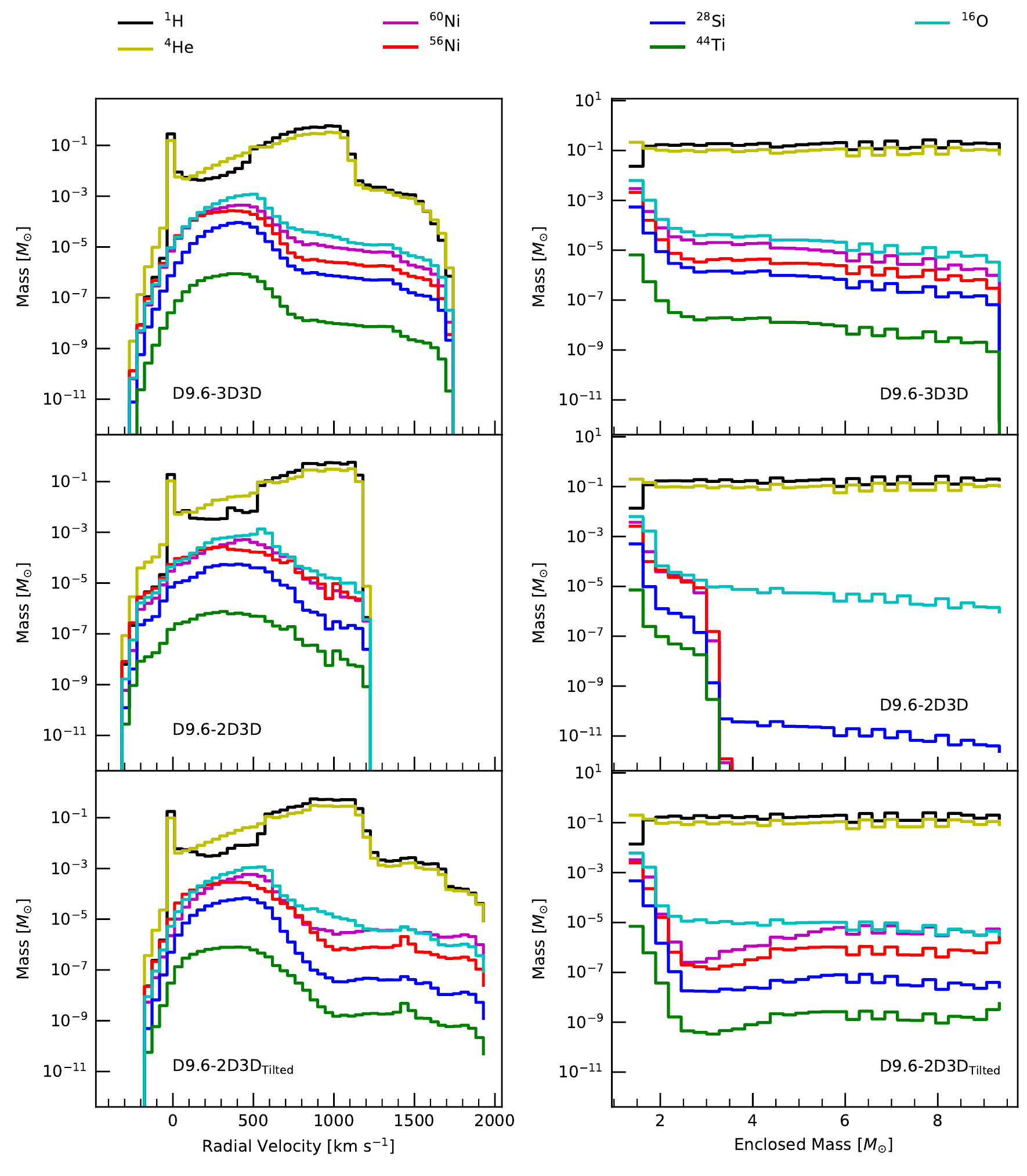}
		\caption{Mass yields of key isotopes binned across radial velocity (left column, 50 bins) and enclosed mass (right column, 30 bins) for each D9.6 model. Note, each bin is consistent across all models for both columns.} \label{fig:d96yields}
	\end{figure*}

	\begin{figure*}
		\includegraphics[width=2.1\columnwidth,clip]{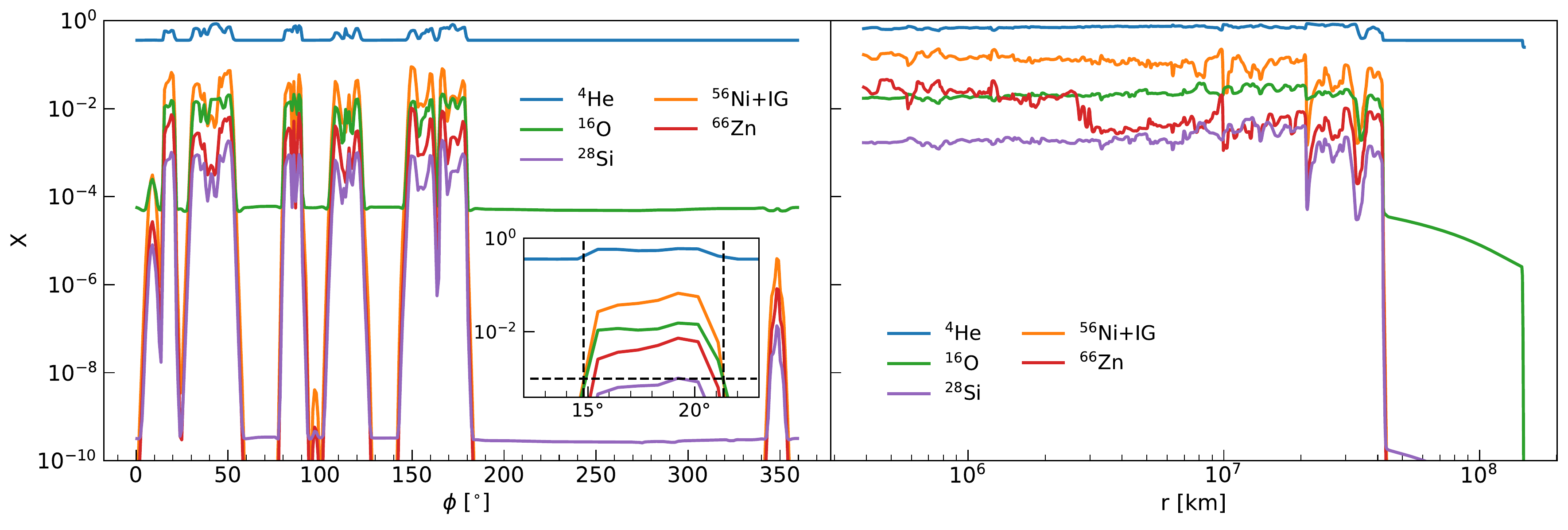}
		\caption{\label{fig:d96plume_frac}
			Left: Mass-fraction vs. azimuth at 10000~s in the x-y-plane ($\theta = 90\degr$) at a constant radius of \ten{4.0}{7}~km. Inset: A magnified region of the 18\degr\ R-T plume. The horizontal dashed line marks a value of $X = 0.1\%$, and the vertical dashed lines mark the angular positions of the 40\% $X_{\isotope{He}{4}}$ external coating. Right: Mass-fraction vs. radius along $\phi = 18\degr$ in the x-y-plane at 10000~s. Note the ``head'' of the plume is located between \ten{3.5}{7}~km~$\leq r \leq$~\ten{4.1}{7}~km and the falloff of \isotope{O}{16} beyond this point is within the progenitor, not the R-T plume.
		}
	\end{figure*}
	
	The shock continues to propagate through the H envelope until hitting the edge of the star, and grid (\ten{1.5}{8}~km), at $\sim$70000~s (19.4 hours).
	We ``rewind'' and declare the end of our simulations at $\sim$62000s, as this is the time where the shock enters the region of the progenitor where the Helmholtz EoS' assumption of fully-ionized hydrogen is no longer valid ($T \lesssim 10000$ K).
	As the shock has only been backtracked to $\sim$\ten{1.4}{8}~km from \ten{1.5}{8}~km, the changes to ejecta morphology, yields, and speeds are negligible.
	Carrying the models further would require accounting for the circumstellar environment and radiation hydrodynamics of shock breakout.
	
	The synchronous conversion between kinetic and internal energy through the entire evolution of the explosion can be seen in Figure~\ref{fig:d96energy}. 
	Both quantities respond to changes in $v_{\rm Shock}$ as the shock front moves through the density structure of the star.
	At the end of our simulations, the internal energy is still in the process of converting to kinetic energy, which starts to converge toward the total energy of the system.
	
	Though the D9.6--\threed\ model clearly has significant extended plume features present at shock breakout (see Figures~\ref{fig:d96velr} and \ref{fig:d96nini}), the majority of trailing R-T bullets have only made it to approximately \ten{1.0}{8}~km and are therefore well short of the surface of the star.
	The early development of features within the He shell, in combination with a smaller relative velocity gap between the shock and fastest moving Ni bullets, enables the further spawning of large R-T mushrooms at the He/H interface.
	This model demonstrates how the early-time asymmetries can impact the late-time evolution of a CCSN.
	Asymmetries at the time \tmap\ seed the initial instabilities from the (C+O)/He interface that spawn further R-T plumes upon reaching the He/H interface --- provided they are moving fast enough relative to the shock.
	This, in turn, affects the efficiency of radial mixing in the outer envelope.
	As seen in the upper panels of Figure~\ref{fig:d96yields}, the instabilities spawned at the He/H interface in this model drive the mixing of metal-rich ejecta beyond the inner 4~\msun\ to the edge of the star.
	The bulk of the bullets end up with a peak centered around 500~\kms\ with the yields for \isotope{O}{16}, \isotope{Si}{28}, \isotope{Ni}{56}, and \isotope{Ni}{60} in that region all in the range between \ten{1}{-3} and \ten{1}{-4}~\msun.  
	However, the extent of radial mixing is quite apparent with a high-velocity tail reaching to $\sim$1750~\kms\ including yields between \ten{1}{-5} and \ten{1}{-7}~\msun\ for these same species.
	
	These bullets are heavily coated in \isotope{He}{4}, with the maximum bins of Figure~\ref{fig:d96yields} exceeding \ten{1}{-1}~\msun\ around 1000~\kms\ and roughly \ten{1}{-3}~\msun\ around 1500~\kms. 
	The most unusual aspect is the internal anatomy of these metal-rich clumps.
	The typical isotope associated with these types of bullets in the literature has consistently been \isotope{Ni}{56}, however \isotope{Ni}{60} seems to fill that role in this star.
	We suspect this is due to the nature of very low-mass CCSNe, but additional simulations with realistic networks are required (to be discussed in E.J. Lentz et al., in prep.).
	Across mass and velocity spaces, \isotope{Ni}{60} is the most abundant of our isotopes in the iron group, and it occupies more of the large-scale features whereas the \isotope{Ni}{56} resides more in the microstructure (Figure~\ref{fig:d96nini}, right).
	Although surprising, the distribution of these isotopes at the time of \tmap\ from \chimera\ (Figure~\ref{fig:compo}, left) makes this the most logical outcome.
	The explosion is surrounded by a shell of \isotope{C}{12}, \isotope{O}{16}, and \isotope{Si}{28}, but the two relevant Ni isotopes are distributed in such a way that the \isotope{Ni}{56} occupies the innermost ejecta whereas the \isotope{Ni}{60} is more extended (Figure~\ref{fig:d96nini}, left).
	The extended \isotope{Ni}{60} features present at the time of \tmap\ grow into further extended structures as the explosion progresses, thus mixing more effectively in mass and velocity in the explosion.
	
	Figure~\ref{fig:d96plume_frac} provides a more detailed look at the compositional structure of the bullets. 
	In the left panel of Figure~\ref{fig:d96plume_frac}, we plot the angular distribution of the composition of R-T plumes residing in the x-y-plane ($\theta = 90\degr$) at a constant radius of \ten{4.0}{7}~km --- marked as the dashed circle slicing the plumes at this radius in Figure~\ref{fig:d96rhoslices}(h).
	The right panel of Figure~\ref{fig:d96plume_frac} displays the composition versus radius of a specific plume residing at $\phi = 18\degr$ in the x-y-plane.
	We choose this time (10000~s) for this inspection because the plumes are much more distinct in Figure~\ref{fig:d96rhoslices}(h) compared to their later appearance and the R-T plumes are simply expanding beyond this point in their evolution.
	In this plane, there exists six extended R-T plumes residing at 18\degr, 43\degr, 86\degr, 115\degr, 155\degr, and 171\degr.
	Two additional, less extended, R-T plumes can also be seen at 9\degr\ and 350\degr.
	All of the extended bullets are consistent in composition, having a \isotope{He}{4} coating that surrounds a metal-rich interior dominated by \isotope{Ni}{56}+IG.
	We find that in addition to the significant amount of \isotope{Ni}{56}+IG present in a single bullet ($X \simeq 0.07$) there also exists a substantial amount of \isotope{O}{16} ($X \simeq 0.02$) and lesser amounts of \isotope{Zn}{66} ($X \simeq 0.008$) and \isotope{Si}{28} present ($X \simeq 0.001$).
	The relatively large presence of \isotope{Zn}{66} in this model is representative of the enhanced $\alpha$-rich, neutron-rich ejecta seen at the end of the \chimera\ run.
	(\isotope{Zn}{66} is the neutron-rich upper limit of the sn160 network.)
	A closer inspection of the 18\degr\ bullet can be seen in the inset in the left panel of Figure~\ref{fig:d96plume_frac}.
	The vertical dashed lines indicate the angular positions of the edge of the 40\% \isotope{He}{4} coating, and the intersection with the horizontal dashed line indicates what value this represents in $X_{\rm \isotope{Ni}{56}+IG}$ (0.001 or 0.1\%), demonstrating the correspondence between these isosurfaces.
	We will track the cocoon of \isotope{He}{4} that encases the heavy-element bullets in Section~\ref{d96tilt} by creating an isosurface at 0.1\% $X_{\rm \isotope{Ni}{56}+IG}$ instead of 40\% \isotope{He}{4}, which allows us to analyze the evolution of the external coating without having additional noise at early-times from the He shell.
	
	\subsection{D9.6-2D3D} \label{d96spin}
\begin{figure}
		\includegraphics[width=\columnwidth,clip]{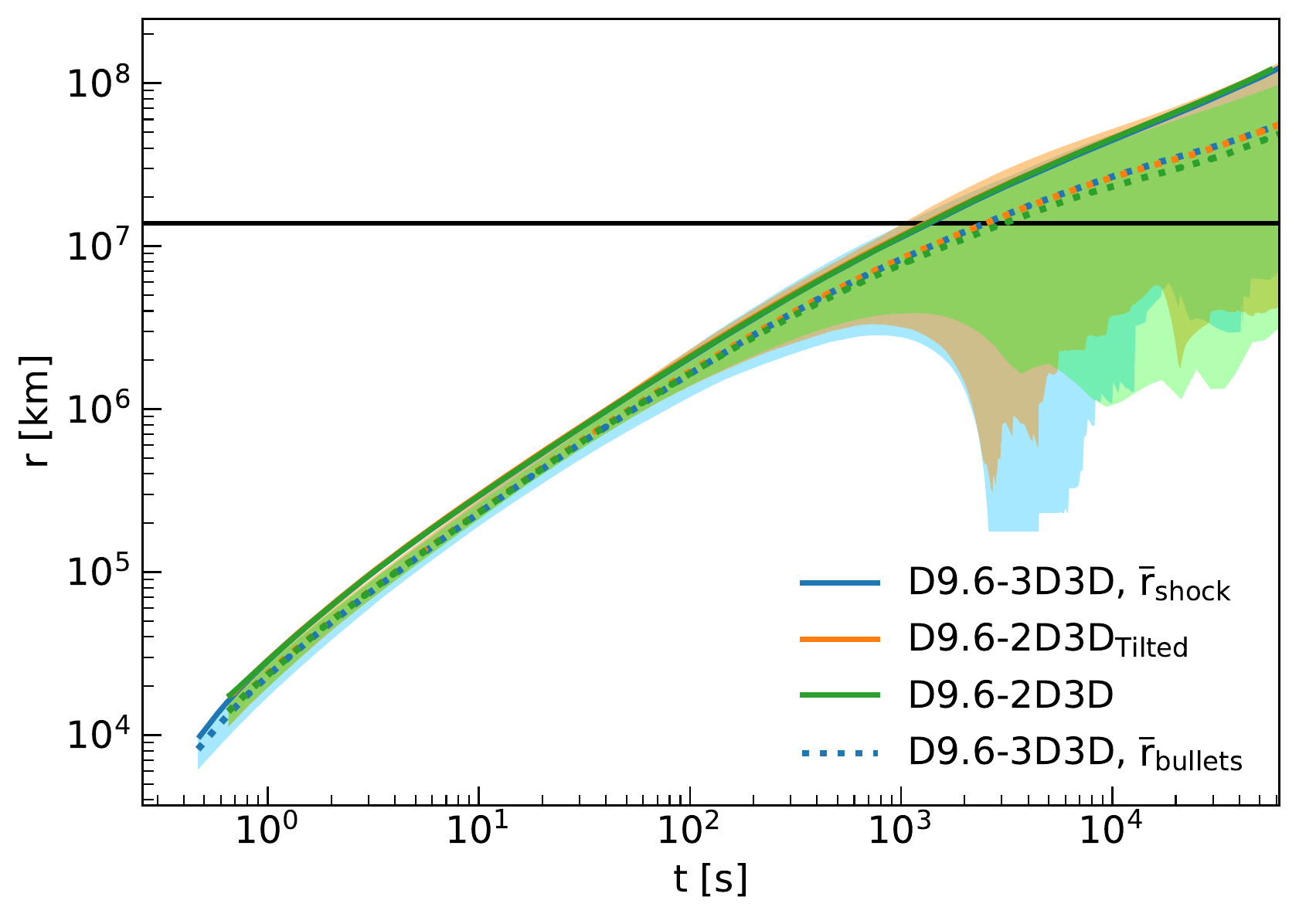}
		\caption{\label{fig:d96shocktime}
			Angle-averaged shock radius (colored, solid lines) and angle-averaged bullet radius of the $X_{\rm \isotope{Ni}{56}+IG} = 3\% $ isosurface (colored, dashed lines) as functions of time. Matching overlaid colored regions highlight the range of $r_{\rm min}$ to $r_{\rm max}$ of a model's respective bullet isosurface. Horizontal black line marks the radius of the He/H interface.
		}
	\end{figure}

	The D9.6--\spun\ model exhibits similar behavior when it comes to the general progression of the shock front, yet differences can be seen when analyzing the leading R-T bullets.
	Although the Ni bullets are able to catch up to the rear of the shock, they are never able to fully interact with it in this model due to a sufficiently large gap in the relative velocity between $v_{\rm Shock}$ and $v_{\rm bullets}$. 
	This can be seen explicitly in Figure~\ref{fig:d96shockvel} in the shock and $X_{\rm \isotope{Ni}{56}+IG}$ velocity curves (green lines).
	The bullets are closest to the shock when the shock front hits the He/H interface at $\sim$\ten{1.0}{7}~km and $\sim$1000~s.
	The maximum velocity of the bullets in this model never rises above the average shock velocity in Figure~\ref{fig:d96shockvel}, explaining why the plumes only minimally interact with the shock.
	Furthermore, in contrast to the D9.6--\threed\ and D9.6--\tilt\ models, the maximum radial position of the $X_{\rm \isotope{Ni}{56}+IG}$ isosurface (Figure~\ref{fig:d96shocktime}, upper edge of the green shaded region) always stays just below the curve representing the average position of the shock (green solid line), which highlights the absence of extended features.
	This minimal interaction leads to the scarcity of large-scale structures and asymmetries in D9.6--\spun.
	Despite that D9.6--\spun\ is mapped roughly 200~ms later than D9.6--\threed\ and the explosion energy is $\sim$13\% higher in D9.6--\spun, the enhanced growth rate of the R-T plumes enabled by the 3D initial state allows D9.6--\threed\ to retain higher velocity bullets.
	
	\begin{figure*}
		\includegraphics[width=2.1\columnwidth,clip]{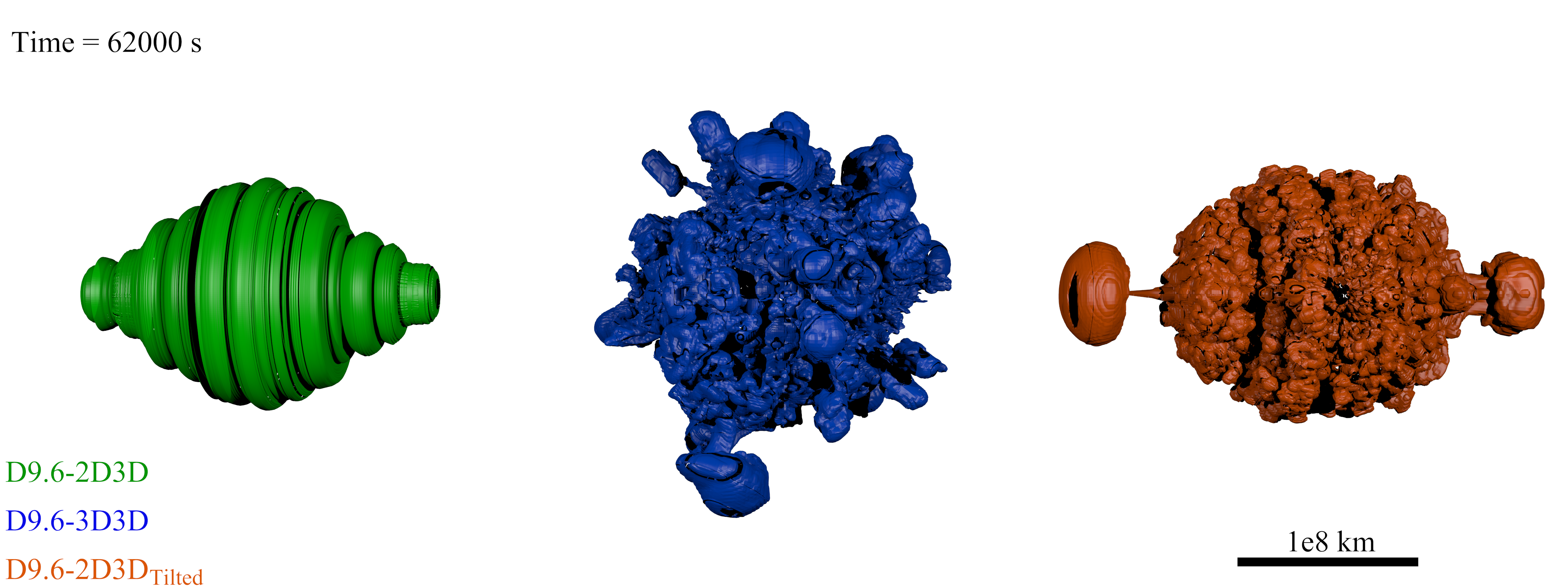}
		\caption{\label{fig:d96sidebyside}
			External coating $X_{\rm \isotope{Ni}{56}+IG} = 0.1\%$ isosurface for the D9.6--\spun\ (green, left), D9.6--\threed\ (blue, center), and D9.6--\tilt\ (orange, right) bullets at shock breakout. Note, the D9.6--\tilt\ isosurface has been realigned in post-processing (i.e. rotated clockwise about its y-axis 90\degr) to match the orientation of the other models.
		}
	\end{figure*}
	
	The lack of macro-structure is apparent in the yields of key isotopes at shock breakout (Figure~\ref{fig:d96yields}).
	The velocity distribution of the ejecta extends only to $\sim$1225~\kms\ in D9.6--\spun, much less than the typical velocities associated with SN1987A, and $\sim$30\% lower in maximum velocity than D9.6--\threed\ (1750~\kms) where the plumes interact with the shock.
	Because the shock is plowing through \isotope{He}{4} and \isotope{H}{1}, the shock can be seen as the ``hump'' in the \isotope{He}{4} curve centered at $\sim$1000~\kms, whereas the bulk of the bullets can be seen as the metal-rich hump further behind, peaking at $\sim$500~\kms.
	The gap between the humps shows how large the relative velocity between the shock front and metal-rich clumps is in D9.6--\spun.
	The distribution gap in velocity further explains the inefficiency of mixing in mass space as well, with most of the metal-rich ejecta only extending to just within 4~\msun.
	Large-scale mushrooms are never spawned from the interaction of the shock with the He/H interface, thus the metal-rich ejecta stays trapped behind the wall of He and is unable to extend its radial mixing.
	
	The shock in D9.6--\spun\ remains roughly spherical, even late in the evolution of the supernova.
	The D9.6--\spun\ model is nearly identical to our D9.6--\twod\ simulations, and this run can be viewed as, in essence, a 2D simulation existing in 3D.
	Without transverse velocities in the 2D \chimera\ model, due to the initially assumed axisymmetry, true 3D behavior never develops (note the unbroken axisymmetry in Figure~\ref{fig:d96sidebyside}, left).
	However, the absence of structure in the shape of the shock is more than made up for by the amount of microstructure present in the inner regions containing the bulk of the R-T instabilities.
	This, in general, is similar to the results of \citet{KiPlJa03} and \citet{WoMuJa15} though with different progenitors.
	From a yields perspective, in both mass and velocity spaces, D9.6--\spun\ is the most similar to the ``\threed'' model presented in \citet{StJaKr20} which uses the same progenitor, referred to as ``z9.6'' (discussed further in Section~\ref{d96compare}).
	
	\begin{figure}
		\includegraphics[width=\columnwidth,clip]{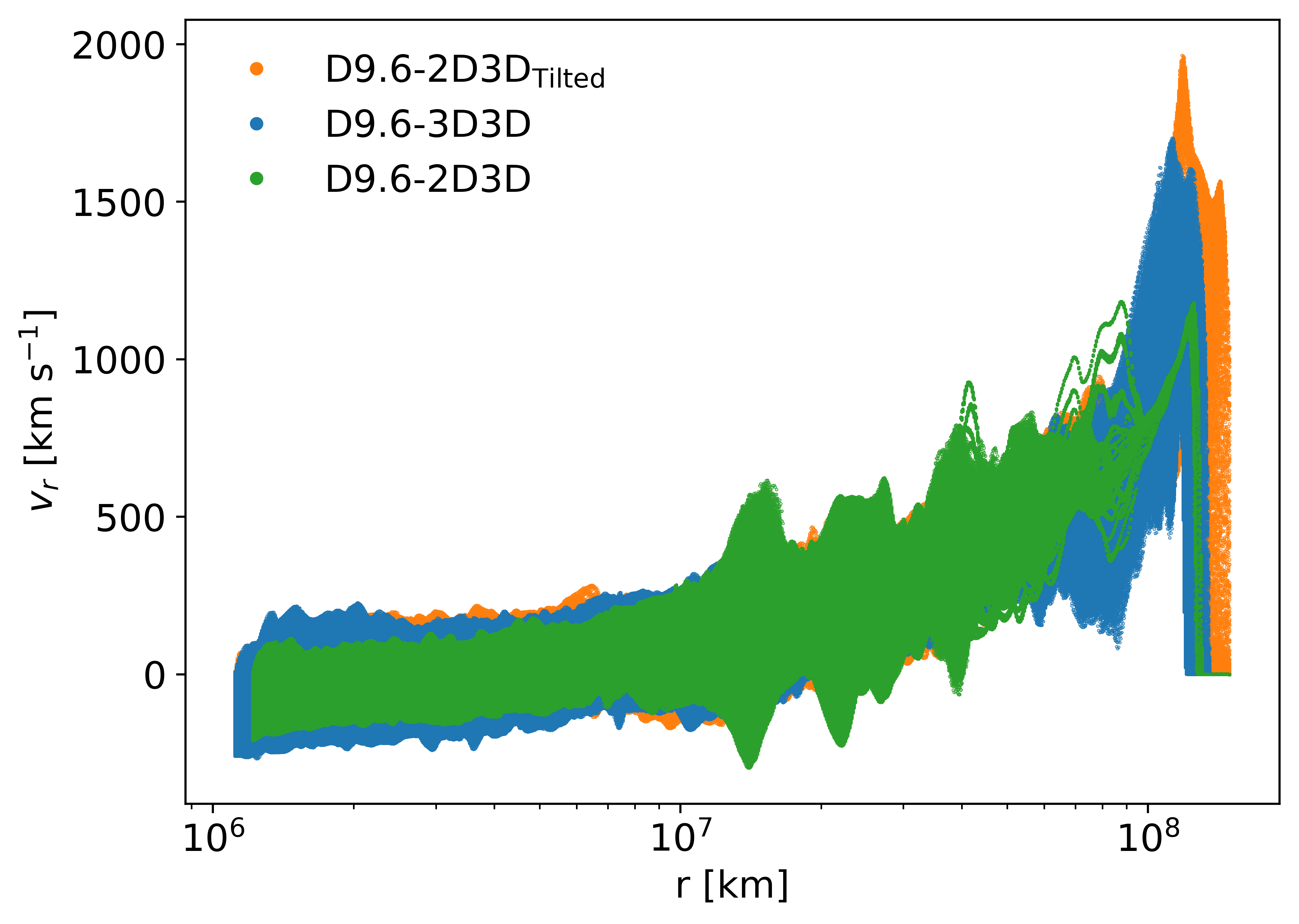}
		\caption{\label{fig:d96scatter}
			Scatter plot of radial velocity versus cell-centered radius for each grid cell at shock breakout for all D9.6 models.
		}
	\end{figure}
	
	\subsection{D9.6-2D3D-Tilted} \label{d96tilt}
	The failure of the D9.6--\spun\ model to depart significantly from its 2D origin is a concrete reminder that the development of fluid flows in response to instabilities is determined by both the strength of the instabilities and the perturbations that seed the fluid flows.
	In the D9.6--\spun\ case, even though mapping to 3D allowed instabilities in the longitudinal direction to develop, the much stronger latitudinal velocity perturbations result in the growth of instabilities that maintain the axisymmetric character of the model. 
	To test this assertion, we constructed the D9.6--\tilt\ model based on a simple coordinate transform.
	Tilting the original 2D \chimera\ model by 90\degr\ on its axis through a coordinate transform introduces longitudinal velocities similar in scale to the previously 100\% latitudinal velocity field.
	Other choices of rotation angle would presumably have similar effects, as long as the misalignment of the old and new axes produces significant longitudinal velocities.
	The presence of both longitudinal and latitudinal velocities seeds the development of features in both coordinate directions.
	Similar to D9.6--\threed, the D9.6--\tilt\ model establishes extended features in its explosion, which allows us to further investigate the morphology of this system.
	Features develop in this version of the ``2D'' model due to the evolution of a spherical-bubble structure rather than a pure toroidal structure imposed by axisymmetry.
	Hence, rotation of the angular velocities enables the explosion to deviate from the initial toroidal structure and start developing bubble-type features when forming the R-T bullets.
	Although only demonstrating a slight deviation initially, the bubble structures are able to retain higher velocities due to experiencing a lower drag to buoyant force ratio and deviate further from axisymmetry as the explosion progresses.
	Echoes of the axisymmetric origin persist, but they do not dominate the entire model to the extent seen in the D9.6--\spun\ model.
	Side-by-side comparisons of all D9.6 models shown in Figure~\ref{fig:d96sidebyside} demonstrate the true impact that tilting the model has on the resulting morphology of the explosion.	
	Clearly, the D9.6--\tilt\ model, though retaining a grossly axisymmetric form, looks more like its true 3D counterpart, while the D9.6--\spun\ model remains almost purely toroidal.
	
	The similarities between D9.6--\tilt\ and D9.6--\threed\ are also apparent in the distribution of radial velocity across the entire grid (Figure~\ref{fig:d96scatter}).
	Unlike D9.6--\spun, both D9.6--\tilt\ and D9.6--\threed\ have a significant number of grid cells occupying high-velocity space beyond a radius of \ten{7.0}{7}~km.
	Additionally, the overall shape of the D9.6--\tilt\ distribution at larger radii looks similar to the D9.6--\threed\ distribution, with a high peak of grid cells before the shock front resulting from the extended features produced in those models.
	Although obscured at lower radii by the D9.6--\spun\ data, D9.6--\tilt\ and D9.6--\threed\ are still consistent, where D9.6--\spun\ is an outlier.
	
	\begin{figure}
		\includegraphics[width=\columnwidth]{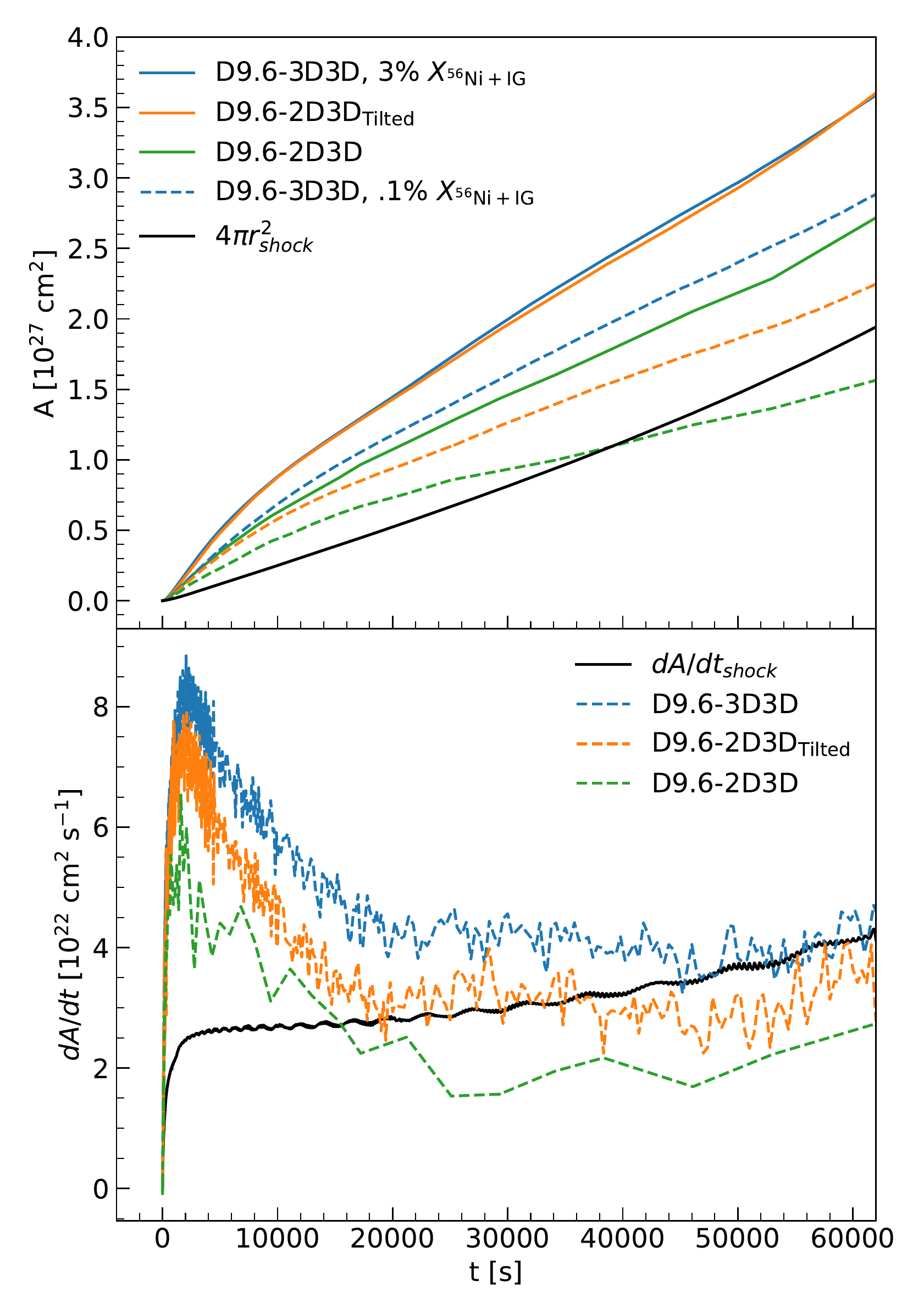}
		\caption{Top: Surface area of the $X_{\rm \isotope{Ni}{56}+IG} = 3\% $ (colored, solid) and $X_{\rm \isotope{Ni}{56}+IG} = 0.1\% $ (colored, dashed) isosurfaces for each D9.6 model. The average shock radii over time across all models are nearly identical, thus only the surface area of the D9.6--\threed\ shock (black, solid) is included. Bottom: Numerical time derivatives of the surface area for the shock (black, solid) and $X_{\rm \isotope{Ni}{56}+IG} = 0.1\% $ (colored, dashed) curves of the top plot. Note that the difference in file output in the D9.6--\spun\ simulation has led to a less dense distribution of data points.} \label{fig:d96area}
	\end{figure}
	
	Analyzing the 3D surface area of the $X_{\rm \isotope{Ni}{56}+IG}$ isosurface shown in Figure~\ref{fig:d96area} further illustrates the divergence between the D9.6--\tilt\ (orange lines) and D9.6--\spun\ (green lines) models.
	The surface area representing the inner anatomy of the bullets (the 3\% isosurface) is nearly identical for D9.6--\threed\ and D9.6--\tilt, while D9.6--\spun\ quickly falls behind in the development of surface area.
	The divergence starts when the shock front encounters the He/H interface, because, once encountering this region, the D9.6--\spun\ model does not have extended features penetrating the interface, which would significantly contribute to the surface area, while the other two models do.
	
	The external coating of the bullets (represented by the 0.1\% isosurface) is visualized in Figure~\ref{fig:d96sidebyside} and displayed more quantitatively by the respective surface area curves in Figure~\ref{fig:d96area} (dashed lines).
	Despite the formation of extended structures in D9.6--\tilt, those features do not occupy as much overall surface area as in the D9.6--\threed\ model, which can be seen in the isosurface plot.
	The biggest plumes in D9.6--\tilt\ do not grow as large as the biggest plumes in D9.6--\threed, which affects the rate of change of the surface area.
	The largest contribution to the surface area occurs at the peak in the bottom plot of Figure~\ref{fig:d96area}, which represents the time that the bullets and shock hit the He/H interface.
	The addition of the He/H mass-shell to the coating of the bullets provides this boost due to the significant amount of extra volume it adds to the bullets.
	As the bullets expand, their surface area grows, and reaches a point near 10000~s where the slopes converge toward the contribution provided by shock expansion.
	The D9.6--\spun\ model converges much faster, as it has no large plumes contributing to its evolution, whereas the other two models are able to stay above the shock expansion curve for longer.
	Although overall converging toward the contributions from the shock, the D9.6--\tilt\ model is able to achieve shock breakout while the total surface area resides above the curve represented by the shock (top panel of Figure~\ref{fig:d96area}).
	This does not occur in D9.6--\spun, as the total surface area is dominated by the shock starting at $\sim$40000~s while almost all of the bullets are trapped behind the He/H mass shell and constantly outpaced by the shock.
	
	Although the external coating isosurface tracks the larger structures of each model, its surface area contribution is not as large as the 3\% isosurface.
	This is due to the fact that, while the 0.1\% tracks larger structures that produce overall greater individual contributions, the amount of smaller individual contributions from the 3\% isosurface is more numerous and builds up to occupy more of the volume, thus representing a larger total surface area.
	
	The overall radial progression of the shock for all three models is nearly identical whether viewed as average shock radius relative to velocity (Figure~\ref{fig:d96shockvel}) or time (Figure~\ref{fig:d96shocktime}).
	Similar to the D9.6--\threed\ model, D9.6--\tilt\ develops its larger scale features in the middle of the He shell near \ten{1.0}{6}~km.
	This can be seen explicitly in Figure~\ref{fig:d96shockvel}, as the maximum velocity of the bullets surpasses the average shock velocity.
	The same type of variations in the velocity profile after \ten{1.0}{6}~km occur in the D9.6--\tilt\ model as they did in D9.6--\threed\ before the velocity of the bullets steadily decline until shock breakout is achieved.
	The plume penetration into the shock is seen further in Figure~\ref{fig:d96shocktime}, as the bullets' maximum radial extent in the D9.6--\tilt\ model (orange shaded region) surpasses the average shock radius and even exceeds that of the equivalent highlighted range in the D9.6--\threed\ model. 
	The average radius of the $X_{\rm \isotope{Ni}{56}+IG}$ clumps in the D9.6--\tilt\ model is nearly identical to that of D9.6--\threed, while the D9.6--\spun\ model deviates around 60~s --- the time when D9.6--\tilt\ and D9.6--\threed\ form their large-scale structures.
	
	\begin{deluxetable}{lccc}
		\tablecaption{Total D9.6 Yields At Shock Breakout \label{tab:yields}}
		\tablecolumns{4}
		\tablewidth{0pt}
		\tablehead{
			\colhead{Species}\hspace{.7cm} & \colhead{\spun} & \colhead{\threed} & \colhead{\tilt} \vspace{-.2cm} \\
			\colhead{ } & \colhead{[\msun]} & \colhead{[\msun]} & \colhead{[\msun]}
		}
		\startdata
		\isotope{H}{1} & 4.995  & 4.958 & 5.017  \\
		\isotope{He}{4} & 3.052 & 3.023  & 3.056   \\
		\isotope{C}{12} & \ten{2.290}{-2} & \ten{2.227}{-2} & \ten{2.226}{-2} \\
		\isotope{O}{16} & \ten{8.125}{-3} & \ten{8.050}{-3} & \ten{7.974}{-3} \\
		\isotope{Si}{28} & \ten{5.118}{-4} & \ten{6.181}{-4} & \ten{5.157}{-4} \\
		\isotope{Ti}{44} & \ten{7.557}{-6} & \ten{7.446}{-6} & \ten{7.705}{-6} \\
		\isotope{Ca}{48} & \ten{1.563}{-4} & \ten{1.419}{-5} & \ten{1.605}{-4} \\
		\isotope{Fe}{52} & \ten{2.777}{-5} & \ten{2.955}{-5} & \ten{2.820}{-5} \\
		\isotope{Ni}{56} & \ten{2.712}{-3} & \ten{2.337}{-3} & \ten{2.767}{-3} \\
		\isotope{Ni}{60} & \ten{4.013}{-3} & \ten{3.669}{-3} & \ten{4.044}{-3} \\
		\isotope{Zn}{66} & \ten{1.383}{-3} & \ten{1.142}{-3} & \ten{1.399}{-3} \\
		Iron Group$_{\rm{NR}}$ & \ten{1.157}{-2} & \ten{1.060}{-2} & \ten{1.174}{-2} \\
		\enddata
		\tablecomments{Iron Group$_{\rm{NR}}$ is defined as all species in our network falling in the range of \isotope{Cr}{49}--\isotope{Ni}{64}, while excluding \isotope{Fe}{52} and \isotope{Ni}{56}. Only cells with a positive radial velocity are considered. This table, with all 160 species, is published in its entirety in the machine-readable format. The species listed above are a selection of the content presented for analysis.}
	\end{deluxetable}
	
	Due to the formation of extended features in D9.6--\tilt, this model bridges the gap between the D9.6--\spun\ and D9.6--\threed\ yields of metal-rich ejecta in both mass and velocity spaces (see Figure~\ref{fig:d96yields}, lower panels).
	Not only is the extent of radial mixing similar to that of D9.6--\threed, but the maximum velocity of the high-velocity tail is 200~\kms\ \emph{larger} than the D9.6--\threed\ model (1950~\kms).
	Once again, the bulk of metal-rich bullets peak at $\sim$500~\kms\ and \ten{1.0}{-4}~\msun.
	The dominant isotope of the iron group in D9.6--\tilt\ is \isotope{Ni}{60}, matching the D9.6--\threed\ model.
	The total yields are relatively consistent across all models (Table~\ref{tab:yields}), with the largest differences arising due to evolution within \chimera\ for the 2D initial condition, while also having a different \tmap\ than the 3D initial condition.
	Additionally, we see relatively low mass loss across all models, $\sim$\ten{7}{-4}~\msun\ lost, where $\sim$1.5\% of this is due to the moving inner boundary (removal of innermost grid cells), and the remaining ejecta lost is due to fallback (matter falling through the inner boundary), most of which is \isotope{He}{4} ($\sim$\ten{3}{-4}~\msun).
	
	\subsection{Comparison to Previous Studies} \label{d96compare}
	Previous works have also studied supernovae from the same progenitor as our D9.6 models, but only one has studied the long-time evolution.
	\citet{StJaKr20} started from the \citet{MeJaMa15} simulation of the neutrino heating phase using similar microphysics in 3D in their version of the 9.6~\msun\ progenitor, but with a smaller nuclear network (15 species + n + p) and lower resolution than our initial state (see Section~\ref{gridnum}).
	They report metal-rich clumps centered around $\sim$300~\kms\ and extending to a maximum of $\sim$500~\kms\ in velocity space, which is clearly slower than our high-velocity tails extending to 1225~\kms, 1750~\kms, and 1950~\kms\ for D9.6--\spun, D9.6--\threed, and D9.6--\tilt, respectively.
	The lower clump velocities in their run also lead to less efficient radial mixing with the metal-rich ejecta only falling within the inner 2~\msun.
	These results are starkly different than the results of our respective D9.6--\threed\ model, which shows mixing to the surface, and are less well mixed than even our simulations initiated from 2D \chimera\ models.
	
	We believe that the notable differences in \citet{StJaKr20} derive from the lower overall diagnostic explosion energy reported at their \tmap, for our initial energy is $\sim$95\% larger in the 3D3D case and even larger in the 2D3D cases.
	Consequently, our shock achieves breakout nearly 12 hours sooner than their reported breakout of $\sim$31 hours, which is an approximately 60\% difference compared to our shock escape before rewinding ($\sim$19.4 hours).
	The weaker overall explosion helps explain why their model does not produce large structures during its evolution, despite it being a 3D model starting with 3D initial conditions.
	The relative velocity gap is too large between the shock front and the leading metal-rich bullets, which enables the R-T plumes to get trapped behind the He/H mass shell as opposed to spawning large features from it (similar to our D9.6--\spun\ model, but to a greater extent).
	This is seen explicitly in Figure 13 from \citet{StJaKr20} (equivalent to our Figure~\ref{fig:d96shockvel}), where their isosurface of $X_{\rm \isotope{Ni}{56}+Tr}$ never reaches maximum velocities that are larger than their average shock velocity.
	Further comparison can be seen in Figure 20 from \citet{StJaKr20} (equivalent to our Figure~\ref{fig:d96velr}), as our bullets look distinctly more elongated while propagating through the He shell.
	
	The total yields at shock breakout for D9.6--\threed\ are shown in Table~\ref{tab:yields}. 
	They differ slightly from the yields at the end of the \chimera\ D9.6-3D model due to the 279~ms of additional burning that occurred. 
	The largest relative changes are in trace abundances of CNO-cycle products (\isotope{O}{14}, \isotope{F}{17}, \isotope{Ne}{18}). 
	More important are enhancements as large as factors of several in species formed during proton-rich, $\alpha$-rich freezeout (e.g., \isotope{Co}{53,54}, \isotope{Cu}{57,59}), including the precursors to the $\nu p$-process (\isotope{Ga}{63,64}, \isotope{Ge}{64}).
	Similar enhancements occur during the same interval in the \chimera\ D9.6-2D model.
	The yields are also broadly comparable to those listed in \citet{StJaKr20}.
	However, differences can be seen in the form of \isotope{Si}{28}, \isotope{Ni}{56}, and the iron group tracer material.
	The amount of \isotope{Si}{28} present at the end of our simulation is 128\% greater than that reported by \citet{StJaKr20}.
	They report approximately 68\% more \isotope{Ni}{56} than our total, which may result from the inclusion of the mass of all iron-group species not included in their network increasing the \isotope{Ni}{56} yield, as they discuss.
	Overall, the amount of neutron-rich iron group material across all of our models is larger by an order of magnitude ($\sim$900\%).
	Considering the amount of \isotope{Ni}{60} present in our model, and that it has essentially ``replaced'' \isotope{Ni}{56} as the traditional bullet material in this simulation, we see comparable or greater total nickel and iron group yields.
	
	Neutron-rich iron peak isotopes are an area where the sn160 network we employed has significant advantage over the smaller network of \citet{StJaKr20}, even with their tracer species.
	Since we see little mass loss of the ejecta during our extended FLASH runs and no significant formation of iron group nuclei during our short period of nuclear burning, the main cause of the discrepancy in the yields seen at shock breakout between D9.6--\threed\ and \citet{StJaKr20} must be how the species were evolved in the \chimera\ and \vertex\ portions of the runs.
	As \citet{StJaKr20} also initiated their shock breakout run from an early-time CCSN simulation \citep{MeJaMa15}, and because there is no discussion of notable mass loss during their late-time evolution, we stress how critical the initial conditions are in this yields comparison.
	
	To a lesser extent, we believe that our higher resolution also impacts the morphology of the system.
	As we will discuss further in Section~\ref{d10}, resolution directly affects the number of R-T plumes spawned when a mass shell fragments.
	The fragmenting phenomenon determines how the shock front is able to be reshaped by the metal-rich bullets.
	A greater number of extant R-T plumes allows for more shock interaction across the entire domain, directly affecting the development of large-scale features.
	However, a more extensive resolution study is required to support this supposition.
	
	For a more general comparison of our D9.6--\threed\ model, we look to the morphology analysis in \citet{WoMuJa15}, who categorized the late-time metal-rich ejecta into three types:
	(1) small clusters of R-T bullets having the fingerprint of early-time asymmetries as in their 15~\msun\ red supergiants (RSGs); (2) fragmented and squished round features as in their 20~\msun\ blue supergiant (BSG); and (3) long extended fingers as in their two 15~\msun\ BSGs.
	Our RSG D9.6 simulations don't seem to fall \emph{completely} into one of these regimes, but the reasoning outlined by \citet{WoMuJa15} does explain why our models look the way they do.
	In their 15~\msun\ BSG models, the steep rise of $\rho r^{3}$ inside the He layer causes a steady deceleration of the shock front as it propagates, and the acceleration/deceleration at the He/H interface is nearly non-existent.
	This allows the bullets to stay close behind the shock and avoid interaction with any of the reverse shocks.
	This is the type of density profile found in D9.6--\threed, as the metal-rich bullets are able to catch up to the rear of the shock in the middle of the He layer due to higher maximum velocities than the average shock velocity.
	
	Despite the presence of extended features in our simulation, they are not as extreme and distinct as those produced by the 15~\msun\ BSG simulations of \citet{WoMuJa15}.
	Our features look like slightly more extended versions of their clustered RSG fingers.
	These clustered structures are associated with early-time asymmetries, and are clearly visible in Figure~\ref{fig:d96nini}, however, the journey for our clumps is different.
	In \citet{WoMuJa15}, the large gap between the shock and the trailing bullets allows for more momentum to build before they collide with the reverse shock produced by the dramatic deceleration at the He/H interface, which does not occur in our simulations due to the smoother density profile of the D9.6 progenitor.
	The development of the bullet shape is strongly impacted by the interaction of the fingers with reverse shocks, which squash the clumps.
	Because the majority of our R-T plumes spawned ahead of the first reverse shock and out of the first mass shell, they completely avoid any reverse shock interaction.
	(The second reverse shock also forms behind the bullets.)
	
	As important as the dynamics of the reverse shock are, neither \citet{StJaKr20} or \citet{WoMuJa15} discuss the phenomenon of the first reverse shock setting up a point-like rebound blast wave as it approaches the inner boundary as seen in our D9.6 models.
	We suspect that this event is missing due to how they moved their inner boundary.
	As discussed in Section~\ref{grid}, we mimicked \citet{WoMuJa15} in the handling of our inner boundary of the grid, but used a 1\% of shock radius criterion as opposed to their 2\%.
	This means we waited longer to move our boundary, thus allowing more accurate interactions near the center of the grid.
	If the collapsing reverse shock encounters the inner boundary when its radial excision is too large, then the reverse shock does not have the opportunity to set up a point-like blast and instead exits the grid.
	Regardless, the impact of this event on the morphology of the system, and its interaction with the PNS wind, needs to be explored further.
	
	\section{D10 - Results} \label{d10intro}
	In this section, we discuss the general progression of the shock in the D10--\spun\ model (Section~\ref{d10}), with specific analysis and comparisons to D10--\tilt\ residing in Section~\ref{d10tilt}.
	
	\subsection{D10-2D3D} \label{d10}
\begin{figure*}
		\includegraphics[width=509.76pt]{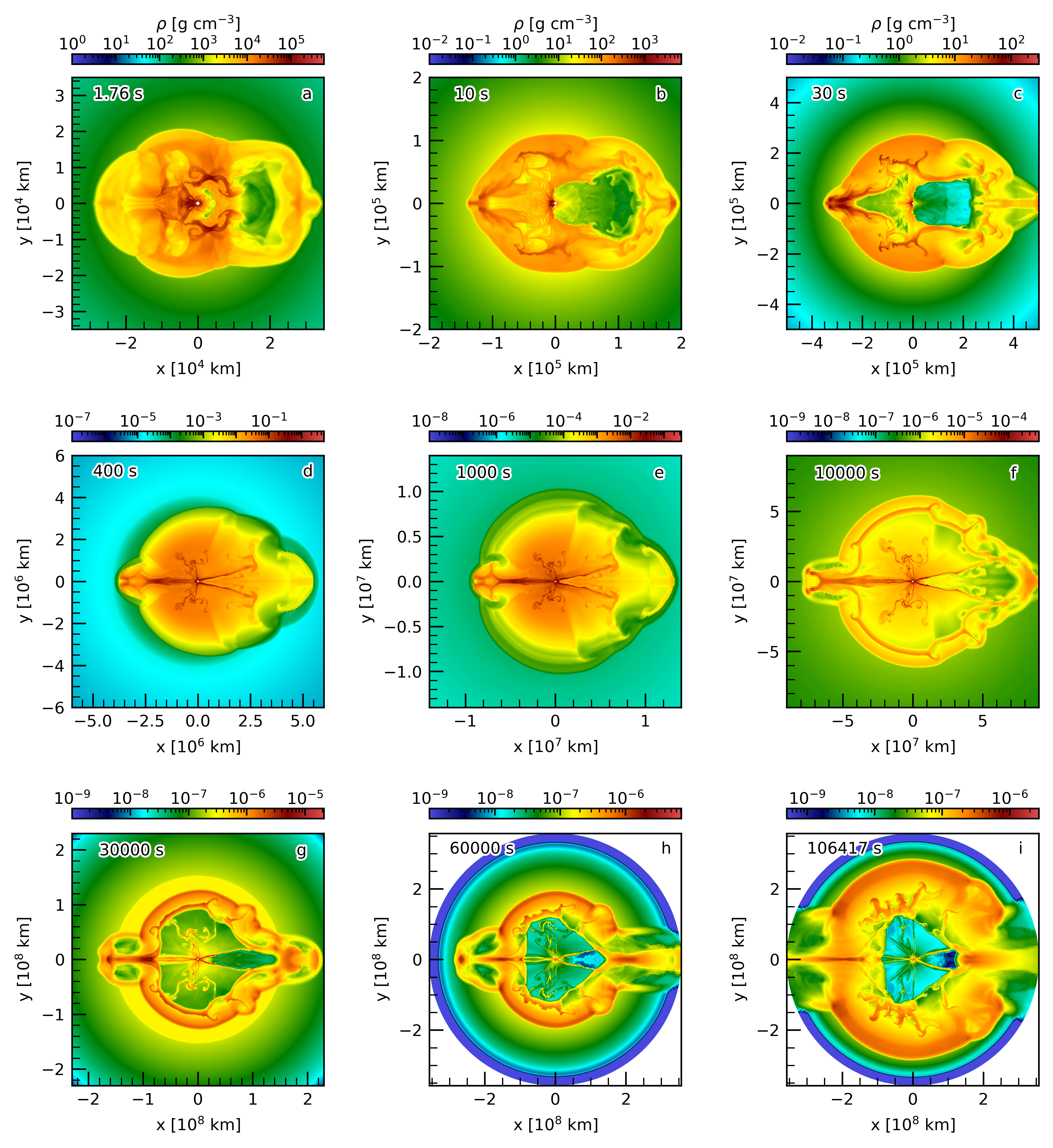}
		\caption{\label{fig:d10rhoslices}
			Slices of density in the D10--\tilt\ model at the displayed times. Note the changes in axis scale and color bar to accommodate the expanding shock. These slices are also consistent with the morphology of the D10--\spun\ model at the given times. The blue to green color discontinuity ahead of the shock in panel (d) represents the position of the He/H interface.
		}
	\end{figure*}

\begin{figure*}
		\includegraphics[width=509.76pt]{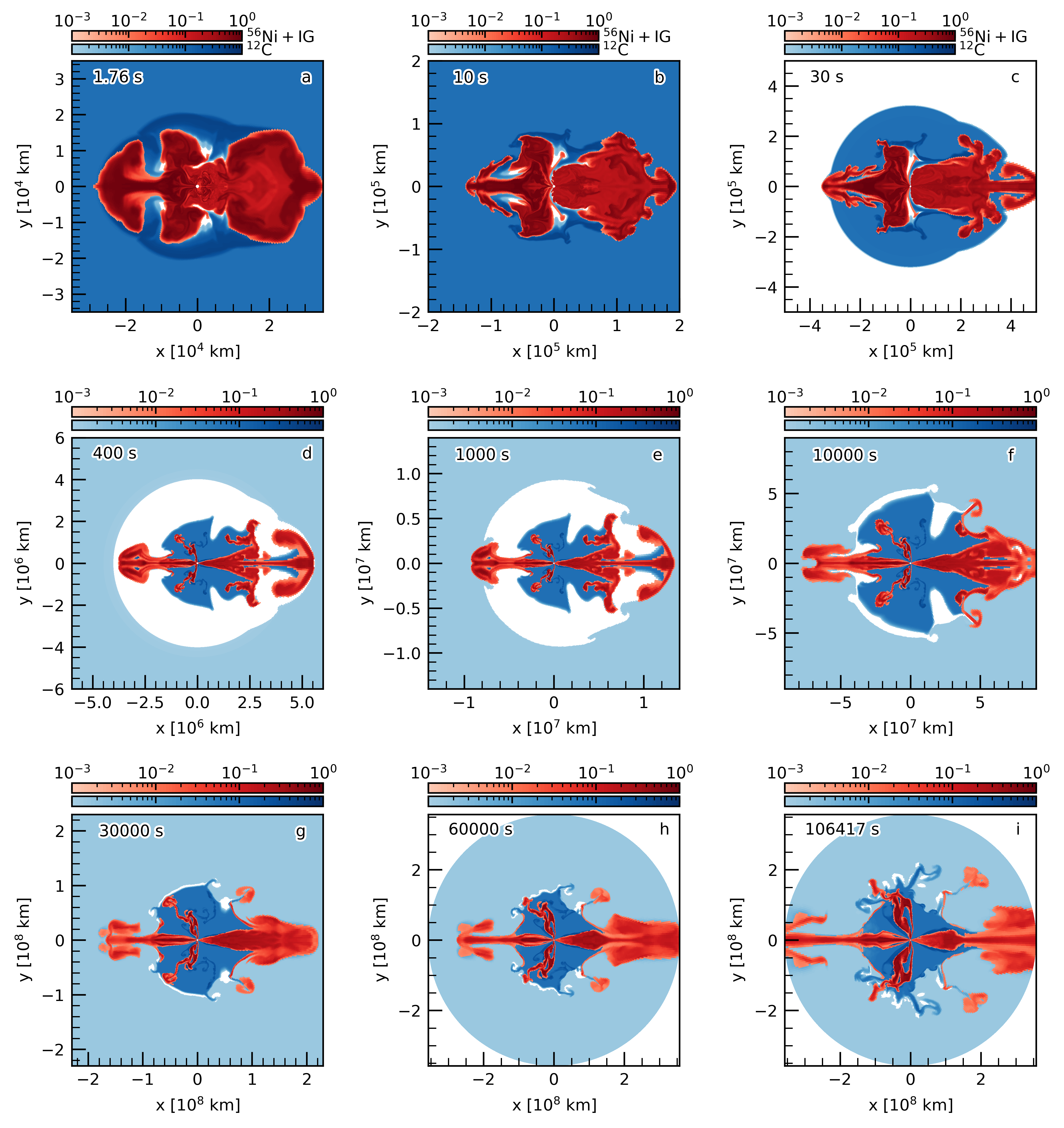}
		\caption{\label{fig:d10bulletslices}
			Slices of $X_{\rm \isotope{Ni}{56}+IG}$ (red) overlaid on $X_{\rm \isotope{C}{12}}$ (blue) in the D10--\tilt\ model at the displayed times (same times as Figure~\ref{fig:d10rhoslices}). These slices are also consistent with the morphology of the D10--\spun\ model at the given times. Note that the different colors help highlight the distribution of ejecta from the various phases of evolution described in Section~\ref{d10}. Red highlights the distribution of phase one, dark blue highlights phase two, and white (material not captured by the limit threshold of either colormap) can be viewed after panel (d) as highlighting the phase three fragmenting shell. 
		}
	\end{figure*}

	Model D10--\spun\ was mapped into FLASH at a much later \tmap\ ($\sim$1.76~s), as nuclear burning ceased much later in this explosion compared to the D9.6 models.
	In addition, the shock front in the D10 is significantly more aspherical from the start (Figures~\ref{fig:d10rhoslices}(a) and \ref{fig:d10bulletslices}(a)) compared to the D9.6 (Figures~\ref{fig:d96rhoslices}(a) and \ref{fig:d96bulletslices}(a)).
	This is a common feature in 2D models of iron core-collapse supernovae, because the development of the explosion depends on the development of large plumes that can deliver energy from the regions of most intense neutrino heating to the shock.
	The three big convective plumes have already shaped the shock front at \tmap\, and will eventually create shear Kelvin-Helmholtz (K-H) instabilities at the interface between the dominant plumes later in its evolution.
	At \tmap, the mean shock radius is $\sim$\ten{2.5}{4}~km, just across the (C+O)/He interface, and is now in the former He-burning shell.
	Unlike the D9.6 simulations that have essentially two phases to the shock progression (pre- and post-He/H interface), the D10 progenitor's density profile gives rise to four distinct phases of evolution.
	Fragmentation of the (C+O)/He shell (phase one) is followed by the acceleration of the shock away from carbon encompassed ejecta (phase two), where this material gets injected into the rear of the fragmenting He/H shell after the shock's deceleration (phase three) before slowly expanding to shock breakout (phase four).
	
	Starting with phase one, the shock is still briefly accelerating after crossing the (C+O)/He interface, which creates a large reverse shock from the subsequent deceleration once fully into the He-burning shell.
	This reverse shock is coupled with the location of the mass shell that once marked the (C+O)/He interface, analogous to D9.6's first reverse shock.
	This promptly shreds and shapes the inner ejecta, as it starts to propagate inward in mass and soon in radius.
	The aspherical shock hits the (C+O)/He interface at slightly different times, leaving a fingerprint in the form of nonuniform fragmentation in its wake.
	Four main R-T plumes are quickly spawned (ignoring the poles), which are reminiscent of the dominant plumes that caused the shock to hit this composition interface, unlike the many small R-T plumes seen in D9.6.
	The main R-T plumes are located at approximately 20\degr, 35\degr, 80\degr, and 130\degr\ from the right pole as seen in Figure~\ref{fig:d10rhoslices}(b) and more distinctly in Figure~\ref{fig:d10rhoslices}(c).
	Of this ``phase one'' material, three out of the four plumes are significantly metal-rich (red features in Figure~\ref{fig:d10bulletslices}(c) at 20\degr, 35\degr, and 130\degr), while the last bullet is rich in \isotope{C}{12} (the darker blue plume at 80\degr).
	These features are mirrored due to the axisymmetry of the initial state and we will omit the mirror features from further discussion because they exhibit the same behavior as their original counterparts.
	
\begin{figure}
		\includegraphics[width=\columnwidth,clip]{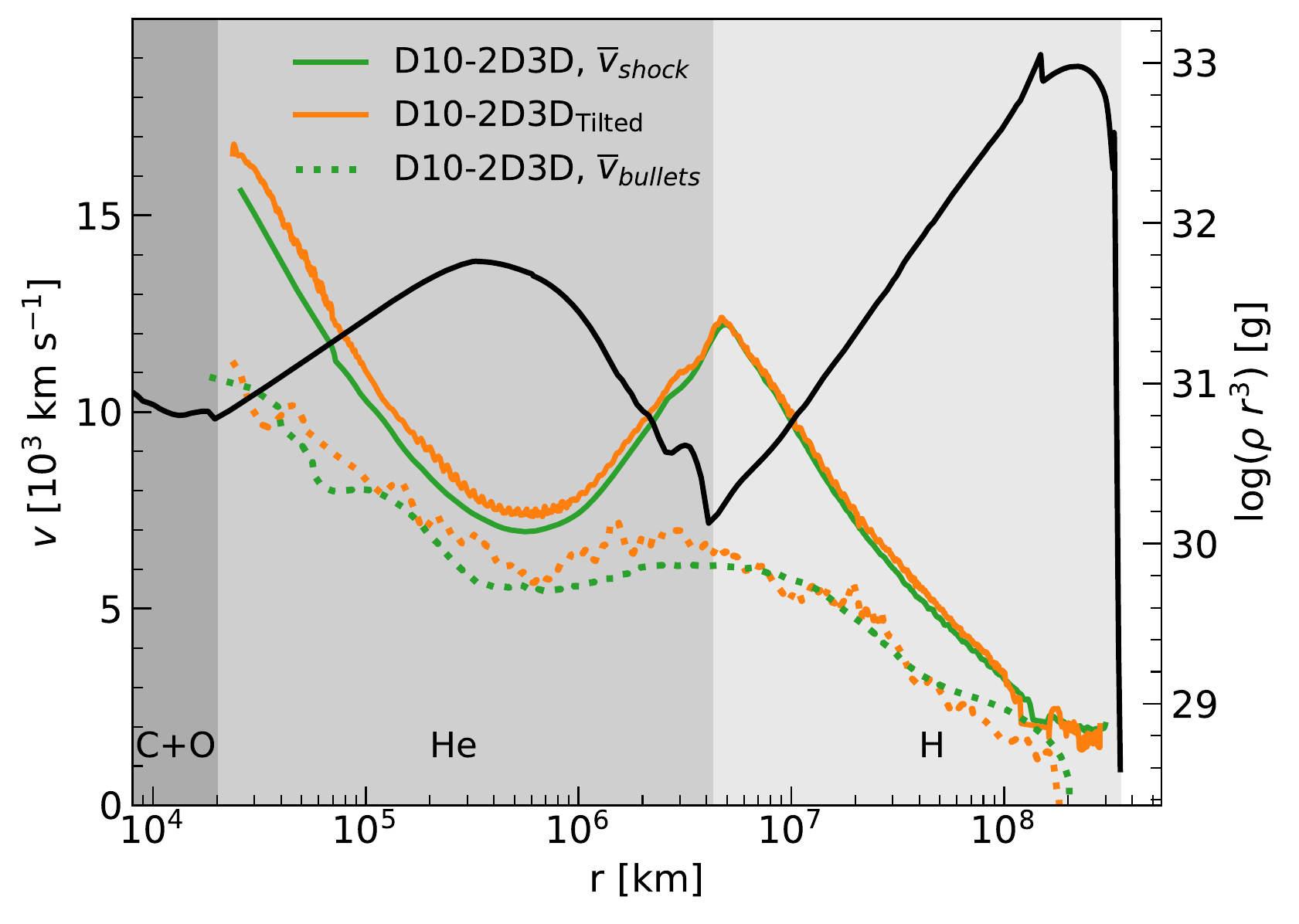}
		\caption{\label{fig:d10shockvel}
			Angle-averaged shock velocity (colored, solid lines) and angle-averaged bullet velocity of the $X_{\rm \isotope{Ni}{56}+IG} = 3\%$ isosurface (colored, dashed lines) for the D10 models as functions of their respective angle-averaged shock or bullet radii. Density profile of the D10 progenitor prior to bounce (black, solid) displays the change of $\rho r^{3}$ and spans the right axis. Grey shaded sections highlight the regions of the (C+O), He, H shells up to the defined interfaces in Table~\ref{tab:prog}.
		}
	\end{figure}

	By 30~s, the shock reaches the density interface at the transition from the He-burning shell to the rest of the He layer (Figures~\ref{fig:d10rhoslices}(c) and \ref{fig:d10bulletslices}(c)).
	Due to the dramatic change in $\rho r^{3}$, the shock encountering this shell starts an acceleration that continues until the shock has fully entered the hydrogen envelope (see change in $v_{\rm Shock}$ starting at $\sim$\ten{3.5}{5}~km in Figure~\ref{fig:d10shockvel}).
	The He-burning shell can be seen throughout the remaining evolution of the explosion as it is propelled forward by the shock.
	The enhanced \isotope{C}{12} from partial He-burning appears as a dark blue carbon ``bubble'' surrounding the inner ejecta in Figure~\ref{fig:d10bulletslices}(c) at $\sim$\ten{3.0}{5}~km while it interacts with the unburned He shell, similar to the helium bubble surrounding the inner ejecta in the D9.6 simulations.
	The unburned He shell in Figure~\ref{fig:d10bulletslices} appears white (lower than the colormap limit of 10$^{-3}$), as it has converted its \isotope{C}{12} to \isotope{N}{14} via the CNO cycle.
	At this point we have entered phase two.
	
	During phase two, until hitting the He/H interface, the shock front significantly outpaces the inner ejecta.
	By 60~s, the four main R-T fingers stretch with extremely thin stems at the base while the reverse shock collapses the material behind them.
	Previously the carbon bubble kept the shape of the shock front, but now the bubble starts to shear at one of the points on its perimeter where the uneven spherical arcs of the shock front hit it prior (at the shock triple point seen earlier at $\sim$50\degr\ in Figures~\ref{fig:d10rhoslices}(c) and \ref{fig:d10bulletslices}(c)).
	This starts to split the bubble and drive a physical wedge between the inner ejecta, which eventually develops into the dramatic dip seen at much later times in the northeastern quadrant of the dark blue bubble in Figures~\ref{fig:d10bulletslices}(d), \ref{fig:d10bulletslices}(e), and \ref{fig:d10bulletslices}(f).
	
	\begin{figure*}
		\includegraphics{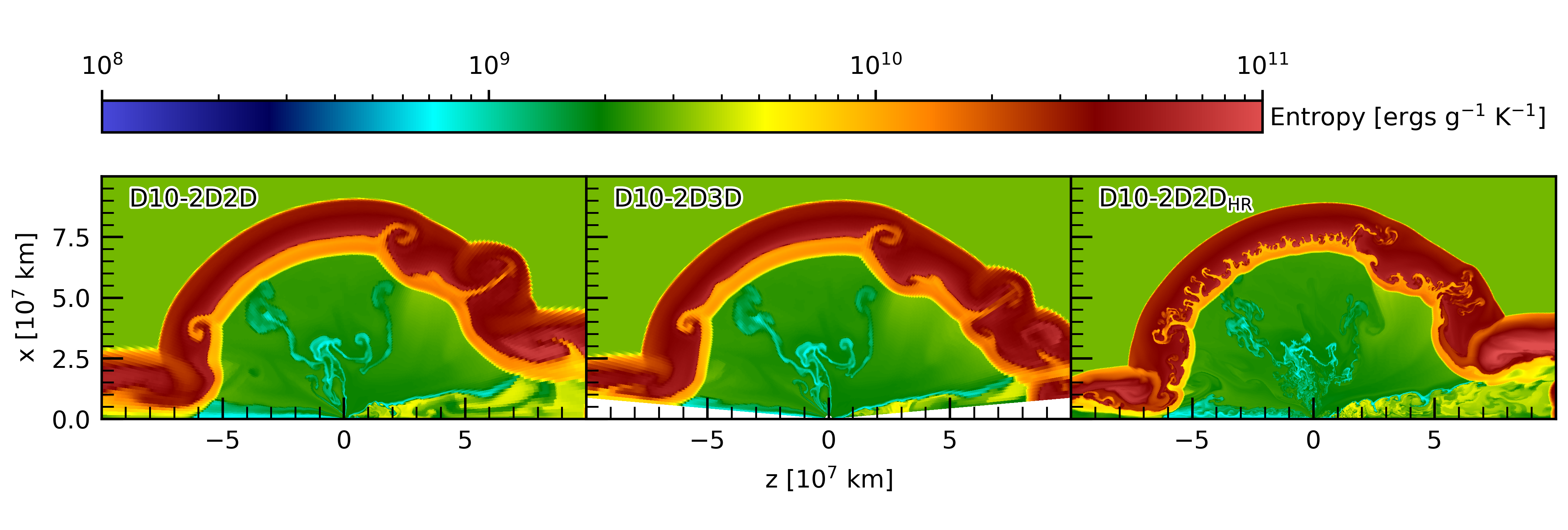}
		\caption{\label{fig:d102dcomp}
			Entropy slice of the D10--\spun\ model (center) compared to 2D simulations of similar resolution (left) and higher resolution (right) at 17500~s.
		}
	\end{figure*}
	
	At $\sim$250~s, the entirety of the first reverse shock has reached the inner boundary of the grid (now at $r=$ \ten{3.4}{4}~km) as we near the end of phase two.
	By this point, the inner ejecta has been completely collapsed by the reverse shock, with all shape and distribution either being huddled close to the inner boundary, or pushed into the four main R-T fingers.
	The first reverse shock approaches the inner boundary significantly more centered about the origin than in the D9.6 models, thus most of it is carried off the grid as opposed to colliding with itself and creating a point-like explosion as seen in the D9.6 models.
	Because of the irregular shape of the first reverse shock, sections of it reach the inner boundary at different times (with the earliest portion reaching the center at $\sim$100~s), which further enables the opportunity to evade collision.
	The only collision that occurs is the portion of the reverse shock produced along the poles that are able to avoid the inner boundary and start to impede the collapsing pressure waves on the opposite side.
	
	Phase three begins at $\sim$400~s (Figures~\ref{fig:d10rhoslices}(d) and \ref{fig:d10bulletslices}(d)) when the main shock encounters the He/H interface and launches a strong reverse shock due to the significant shock deceleration (see in Figure~\ref{fig:d10shockvel} the sharp change in $v_{\rm Shock}$ at $\sim$\ten{4.5}{6}~km).
	The second reverse shock is coupled to the location of the mass shell of the He/H interface, unlike the decoupled second reverse shock in the D9.6 models.
	This shell starts to fragment quickly, with an R-T instability forming promptly at the point where the main shock hit unevenly.
	Since the eastern side of the shock encounters the interface first, this region of the explosion develops its ``phase three'' R-T plumes the quickest, which can be seen explicitly as the two white instabilities in the northeastern quadrant in Figure~\ref{fig:d10bulletslices}(e) at a radius of $\sim$\ten{9.0}{6}~km.
	Additional fragmentation occurs later, forming singular, but dominant, R-T instabilities in succession.
	In our higher resolution 2D tests, we find the development of R-T instabilities to be much more abundant and the fragmentation to be much more uniform.
	
	The deceleration of the shock front allows the trailing phase two and phase one material (the carbon bubble and R-T plumes within this bubble, respectively) to eventually get injected into the rear of the fragmenting He/H shell.
	The carbon bubble achieves this first, as a portion of it first reaches the fragmenting He/H shell and its reverse shock at 3000~s.
	By 10000~s (Figure~\ref{fig:d10bulletslices}(f)), some of the He/H R-T plumes have penetrated the rear of the shock front, and the second reverse shock continues to propagate inward in mass, which allows the remaining regions of the phase two carbon bubble to catch up to it.
	(The dark blue bubble in Figure~\ref{fig:d10bulletslices}(f) catches up to the white.)
	Note that the two metal-rich phase one R-T plumes in the northeast have at this point merged and burrowed through both the carbon bubble and the second reverse shock (see red R-T plume at $\sim$40\degr\ in Figure~\ref{fig:d10bulletslices}(f)).
	At 30000~s, the remaining phase one R-T instabilities reach and interact with this shell as well.
	(The 80\degr\ and 130\degr\ R-T plumes reach the front edges of the dark blue and white at $\sim$$10^8$~km in Figure~\ref{fig:d10bulletslices}(g).)
	Additionally, the fragmenting shell, which was once only composed of helium and hydrogen, is now enriched in the phase two carbon.
	(The blue bubble now occupies the inner anatomy of the previously white R-T plumes in Figure~\ref{fig:d10bulletslices}(g).)
	
	Phase four is the simplest of all our phases, as most features within the explosion are solely expanding radially.
	At about 40000~s, the shock crosses a sudden density spike in the middle of the H shell ($\rho r^{3}$ spike at $\sim$\ten{1.5}{8}~km in Figure~\ref{fig:d10shockvel}).
	This does not produce a third reverse shock, but it does spawn a noticeable pressure wave that starts propagating inward in mass (and eventually in radius), as the shock experiences a jolt seen as fluctuations in its velocity starting at this point (see $v_{\rm Shock}$, solid lines, in Figure~\ref{fig:d10shockvel}).
	Although some of the He/H R-T plumes penetrated the rear of the main shock earlier, they have lost momentum trying to dig their way through the shock and are now being outpaced by it.
	By 60000~s, this model has partial shock breakout at the poles and the shock exits the grid along the pole.
	As these polar flows are artifacts of the assumed symmetry in \chimera, we continue the simulation to determine when the remainder of the shock front would achieve shock breakout.
	From this point forward, we provide analysis on the wedge of data that exclude the polar regions.
	(The wedge considers polar angles $30\degr \leq \theta \leq 150\degr$ across all $\phi$.)
	
	The second reverse shock further collapses the phase two carbon bubble, and the stems of the phase one R-T plumes within it, as it starts to progress inward in radius at $\sim$70000~s.
	This continues until full shock breakout is achieved when the (non-pole) shock leaves the grid (\ten{3.57}{8}~km) at $\sim$140000~s (38.8~hours).
	We rewind the end of our simulation to $\sim$110000~s, when the aforementioned ``wedge'' of the shock enters the region of the progenitor that is partially ionized.
	By this time, the majority of trailing R-T bullets are at $\sim$\ten{2.0}{8}~km, approximately 12~hours behind the shock front.
	
	The D10--\spun\ model keeps its toroidal shape through its entire evolution, like the D9.6--\spun\ model.
	The average velocity of the metal-rich clumps is significantly lower than the average velocity of the shock (see consistent gap between the green curves in Figure~\ref{fig:d10shockvel}).
	The velocity gap between the two increases when the shock front starts to accelerate down the density gradient as it approaches the He/H interface, which enlarges the relative velocity gap to a difference of $\sim$7000~\kms.
	Although this does not allow for any interaction with the main shock, it does allow for the main R-T clumps to grow rather elongated before encountering the He/H mass shell and reverse shock.
	
	Burrowing through the He/H mass shell is what establishes the final morphology of the CCSN, as this greatly shapes the ejecta and has the ability to spawn further R-T plumes.
	However, the fragmentation of this shell is quite minimal, and the perturbation from the trailing R-T clumps only seems to add to its bulk at the point of collision.
	Although some R-T plumes are seeded from this event, the development of the extended structures echos only the previously trailing asymmetries, rather than having a fully fragmented shell across all angles.
	Figure~\ref{fig:d102dcomp} shows how different the environment is between D10--\spun\ and a high-resolution D10--\twod\ model.
	The D10--\spun\ model has three main He/H R-T features forming out of the fragmenting shell as the trailing instabilities catch up to it, while the high resolution D10--\twod\ simulation has numerous R-T plumes developing at the equivalent time.
	
	\begin{figure*}[p]
		\includegraphics{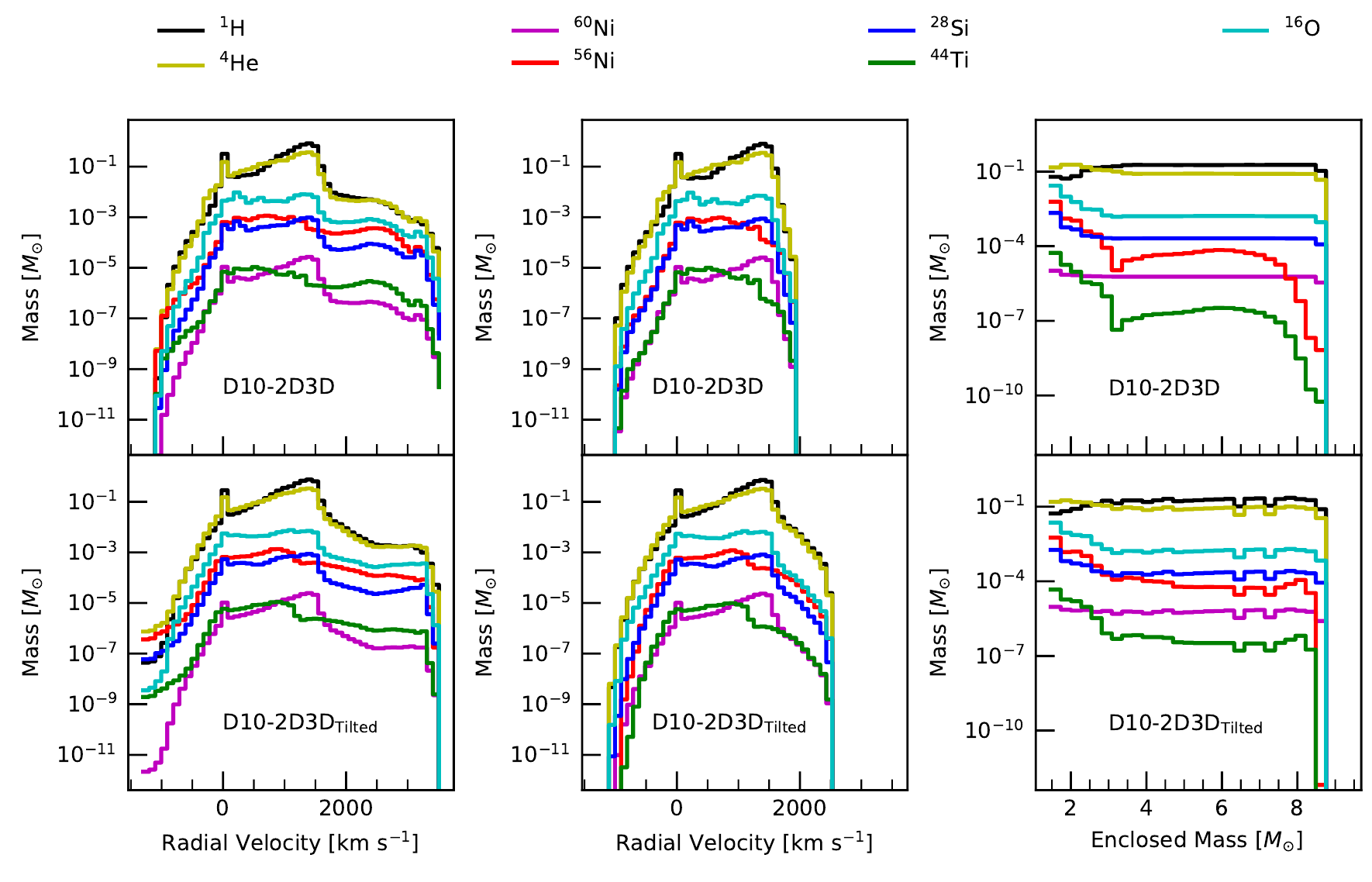}
		\caption{Mass yields of key isotopes binned across radial velocity (left, center columns -- 50 bins) and enclosed mass (right column -- 30 bins) for each D10 model. Note that each bin is consistent across all models for each column, and that both the center and right columns exclude the polar flows via considering a wedge of the data defined in Sections~\ref{d10} and \ref{d10tilt}.} \label{fig:d10yields}
	\end{figure*}
	
	Naturally, the greater number of R-T plumes is not surprising given a much higher resolution, but we provide it here as an example of how the morphology can evolve much differently if the trailing R-T plumes encounter a fragmenting shell equivalent to that of the D10--\twod\ high resolution model.
	While the bullets in the high resolution D10--\twod\ model still have a fingerprint of the clumps that collided with the second reverse shock, there is a much more complex angular distribution of ejecta with much more mixing close to the rear of the shock front.
	This complex environment does not occur in the D10--\spun\ model (or in D10--\tilt\ as we will discuss in Section~\ref{d10tilt}), which shows a morphological environment that echoes the asymmetries of the past.
	The D10--\spun\ model is eerily similar to its D10--\twod\ counterpart of the same resolution (compare center to left panel of Figure~\ref{fig:d102dcomp}).
	As was apparent with the D9.6--\spun\ model, a basic \spun\ mapping does not provide much benefit over running a 2D simulation with similar resolution, due to the absence of longitudinal velocities.
	
	\subsection{D10-2D3D-Tilted} \label{d10tilt}
	The D10--\spun\ model does not seem to accurately portray the long-term evolution of a strongly axisymmetric explosion, due to the lack of initial longitudinal velocities and exaggerated polar flows from an unfortunate interaction between the 2D \chimera\ model's polar flow and the excised cone in FLASH.
	The D10--\tilt\ model alleviates the interaction with the excised cone, though the polar flow itself is still present as it has been tilted fully onto the FLASH grid.
	
	At first glance, the first column of the yields in Figure~\ref{fig:d10yields} do not show much change in the ejecta distribution in velocity space between D10--\spun\ and D10--\tilt.
	The dominance of the poles in both models drowns out contributions from the rest of the ejecta to the higher velocity matter and hides the microstructure in the first column of Figure~\ref{fig:d10yields}.
	Because the poles in the D10 model are so dramatic, this provides a counterexample to the argument that the yields distribution in the D9.6--\tilt\ model could potentially be misleading due to more of the polar flow being present on the grid compared to its respective D9.6--\spun\ model.
	If that were the case, then we would see a more dramatic difference in the distribution of the ejecta when comparing the upper and lower panels of the first column in Figure~\ref{fig:d10yields}.
	Clearly, we do not.
	
	\begin{deluxetable}{lccc}
		\tabletypesize{\small}
		\tablecaption{Total D10 Yields At Shock Breakout \label{tab:d10yields}}
		\tablecolumns{3}
		\tablewidth{0pt}
		\tablehead{
			\colhead{Species}\hspace{.7cm} & \colhead{\spun} & \colhead{\tilt} \vspace{-.2cm} \\
			\colhead{ } & \colhead{[\msun]} & \colhead{[\msun]}
		}
		\startdata
		\isotope{H}{1} & 4.131  & 4.277  \\
		\isotope{He}{4} & 2.488 & 2.533  \\
		\isotope{C}{12} & \ten{4.546}{-2} & \ten{4.580}{-2} \\
		\isotope{O}{16} & \ten{7.985}{-2} & \ten{8.076}{-2} \\
		\isotope{Si}{28} & \ten{7.842}{-3} & \ten{7.948}{-3} \\
		\isotope{Ti}{44} & \ten{8.952}{-5} & \ten{9.481}{-5} \\
		\isotope{Ca}{48} & \ten{1.132}{-6} & \ten{1.164}{-6} \\
		\isotope{Fe}{52} & \ten{2.146}{-4} & \ten{2.207}{-4} \\
		\isotope{Ni}{56} & \ten{9.860}{-3} & \ten{1.115}{-2} \\
		\isotope{Ni}{60} & \ten{1.571}{-4} & \ten{1.616}{-4} \\
		\isotope{Zn}{66} & \ten{6.482}{-6} & \ten{6.572}{-6} \\
		Iron Group$_{\rm{NR}}$ & \ten{1.068}{-2} & \ten{1.101}{-2} \\
		\enddata
		\tablecomments{These yields exclude contributions from the polar flows. Iron Group$_{\rm{NR}}$ is defined as all species in our network falling in the range of \isotope{Cr}{49}--\isotope{Ni}{64}, while excluding \isotope{Fe}{52} and \isotope{Ni}{56}. Only cells with a positive radial velocity are considered. This table, with all 160 species, is published in its entirety in the machine-readable format. The species listed above are a selection of the content presented for analysis.}
	\end{deluxetable}
	
	To reveal microstructure obscured by the poles, we further analyze the yields by considering a wedge of the models that excludes contributions on the grid from the polar flows (second, third columns of Figure~\ref{fig:d10yields}).
	The wedge for D10--\tilt\ is the same wedge discussed in Section~\ref{d10} for the D10--\spun\ model ($30\degr \leq \theta \leq 150\degr$ across all $\phi$), but is applied after a 90\degr\ coordinate transform (i.e. after ``undoing'' the tilt).
	Through this, we actually see more of an effect that tilting the model has provided, as D10--\tilt\ has an apparent higher velocity tail ($\sim$2500~\kms) compared to its D10--\spun\ counterpart ($\sim$1900~\kms) when comparing models in the second column of Figure~\ref{fig:d10yields}.
	Comparing models in mass space (third column of Figure~\ref{fig:d10yields}) shows higher yields for D10--\tilt\ in the outer regions.
	This can clearly be seen in the extent of \isotope{Ni}{56} and \isotope{Ti}{44}, which both drop significantly in the D10--\spun\ model at 7.5~\msun\ (top row, third column of Figure~\ref{fig:d10yields}).
	In contrast, for the D10--\tilt\ model both \isotope{Ni}{56} and \isotope{Ti}{44} extend to 8.5~\msun, joining the lighter elements in the ejecta (bottom row, third column of Figure~\ref{fig:d10yields}).
	The total yields (Table~\ref{tab:d10yields}) further reveal this difference, with roughly 6\% and 13\% greater \isotope{Ti}{44} and \isotope{Ni}{56} yields, respectively, in the D10--\tilt\ model.
	Due to more of the polar flows, which originate from the hot bubble, being included on the grid, these isotopes (plus \isotope{Fe}{52}) are some of the key differences relative to the D10--\spun\ model, while the rest of the yields are relatively consistent between D10 models.
	Although the poles are excised for both models in Table~\ref{tab:d10yields}, the relevant species are more heavily mixed into surrounding areas during the explosion in the D10--\tilt\ model, thus these species are more abundant than for D10--\spun.
	
	\begin{figure}
		\includegraphics[width=\columnwidth,clip]{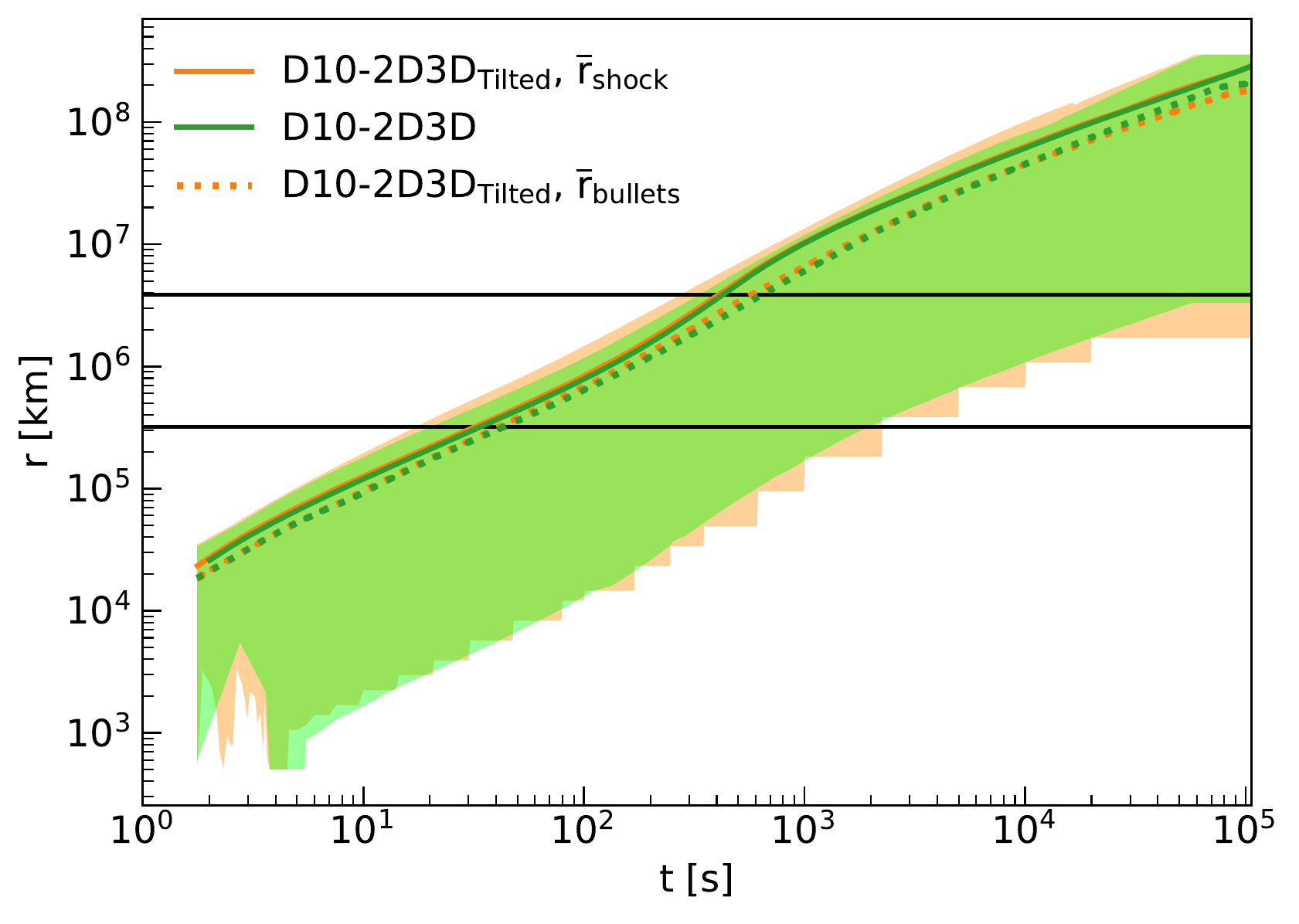}
		\caption{\label{fig:d10shocktime}
			Angle-averaged shock radius (colored, solid lines) and angle-averaged bullet radius of the $X_{\rm \isotope{Ni}{56}+IG} = 3\%$ isosurface (colored, dashed lines) as functions of time. Matching overlaid colored regions highlight the range of $r_{\rm min}$ to $r_{\rm max}$ of a model's respective bullet isosurface. The horizontal black lines mark the radii of the He burning shell to inert He layer transition (bottom line) and He/H composition interface (top line).
		}
	    \vspace{5mm}
	\end{figure}
	
	\begin{figure*}
		\includegraphics[width=2.1\columnwidth,clip]{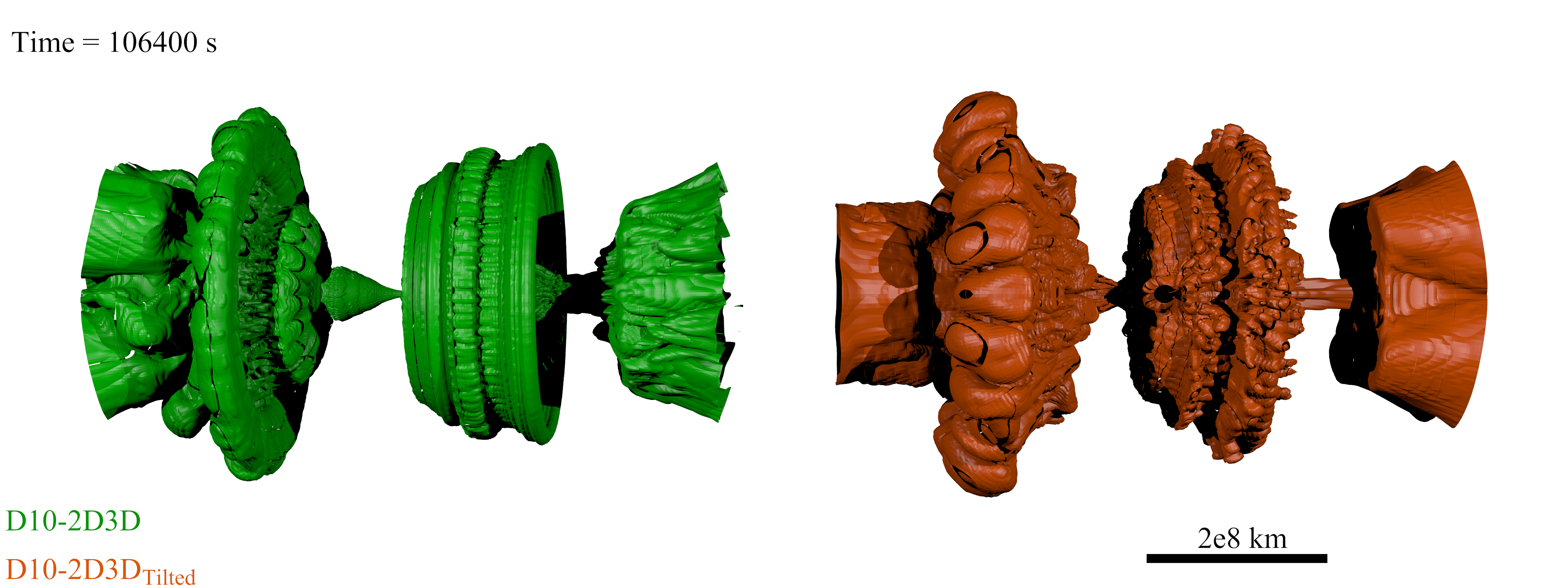}
		\caption{\label{fig:d10sidebyside}
			External coating $X_{\rm \isotope{Ni}{56}+IG} = 0.1\%$ isosurface for the D10--\spun\ (green, left) and D10--\tilt\ (orange, right) bullets at shock breakout. Note, the D10--\tilt\ isosurface has been realigned in post-processing (i.e. rotated clockwise about its y-axis 90\degr) to match the orientation of the other model. The open ended ``caps'' are due to the poles evolving off the grid much earlier in the simulation.
		}
	    \vspace{5mm}
	\end{figure*}
	
	\begin{figure}
		\includegraphics[width=\columnwidth,clip]{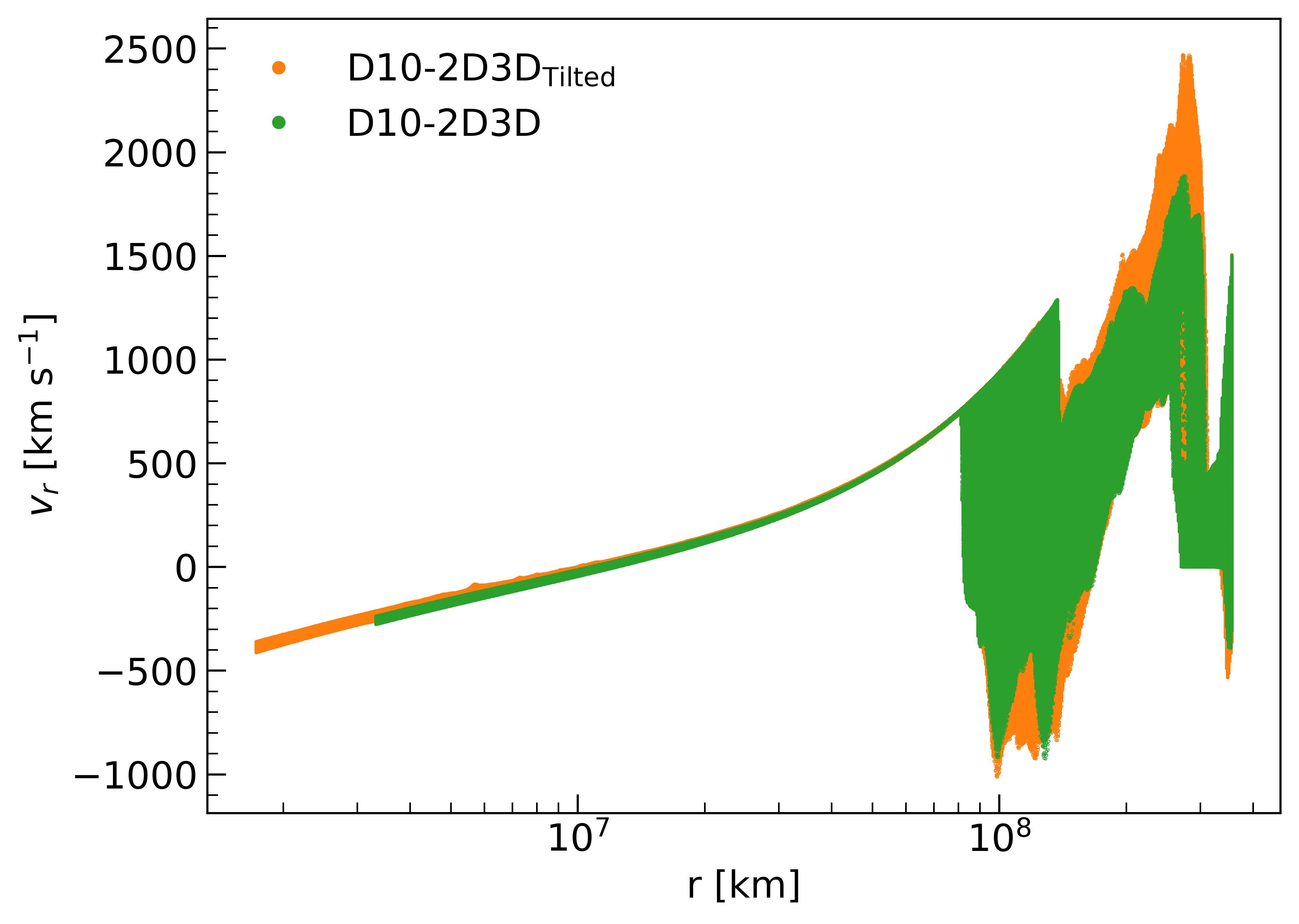}
		\caption{\label{fig:d10scatter}
			Scatter points of a grid cell's radial velocity versus cell-centered radius at shock breakout for each D10 model. Note, cells in the polar flows have been excluded via considering a wedge of the data.
		}
	    \vspace{5mm}
	\end{figure}
	
	The consistency of the ejecta for the two D10 models is matched by the consistency in shock progression (colored, solid curves in Figure~\ref{fig:d10shocktime}).
	Even the average radii of the $X_{\rm \isotope{Ni}{56}+IG}$ isosurfaces (colored, dashed curves) are quite similar.
	Despite this, the D10--\tilt\ model develops more spherical-bubble structures during its evolution (Figure~\ref{fig:d10sidebyside}, right), due to the initial longitudinal and latitudinal velocities.
	This is consistent with what happened in the D9.6--\tilt\ model.
	D10--\tilt\ is slightly less axisymmetric than D10--\spun\ in Figure~\ref{fig:d10sidebyside} and has more structure in its central and outer regions.
	Therefore, these metal-rich clumps in the D10--\tilt\ model retain slightly higher velocities (dashed orange curve in Figure~\ref{fig:d10shockvel}) over its D10--\spun\ counterpart (dashed green curve in Figure~\ref{fig:d10shockvel}) until the He/H interface when the shock starts to decelerate and the second reverse shock forms.
	The second reverse shock dictates the subsequent velocity profile of the clumps, limiting their velocities as they try to burrow through it, bringing the average clump velocities back together as both dashed curves decrease until shock breakout.
	Although the \emph{average} velocities of the clumps for both models obtain similar values near shock breakout, the \emph{overall} velocity distribution across the analysis wedge domain (Figure~\ref{fig:d10scatter}) shows that the D10--\tilt\ model still retains higher velocities in the outer envelope.
	
\begin{figure}
		\includegraphics[width=\columnwidth]{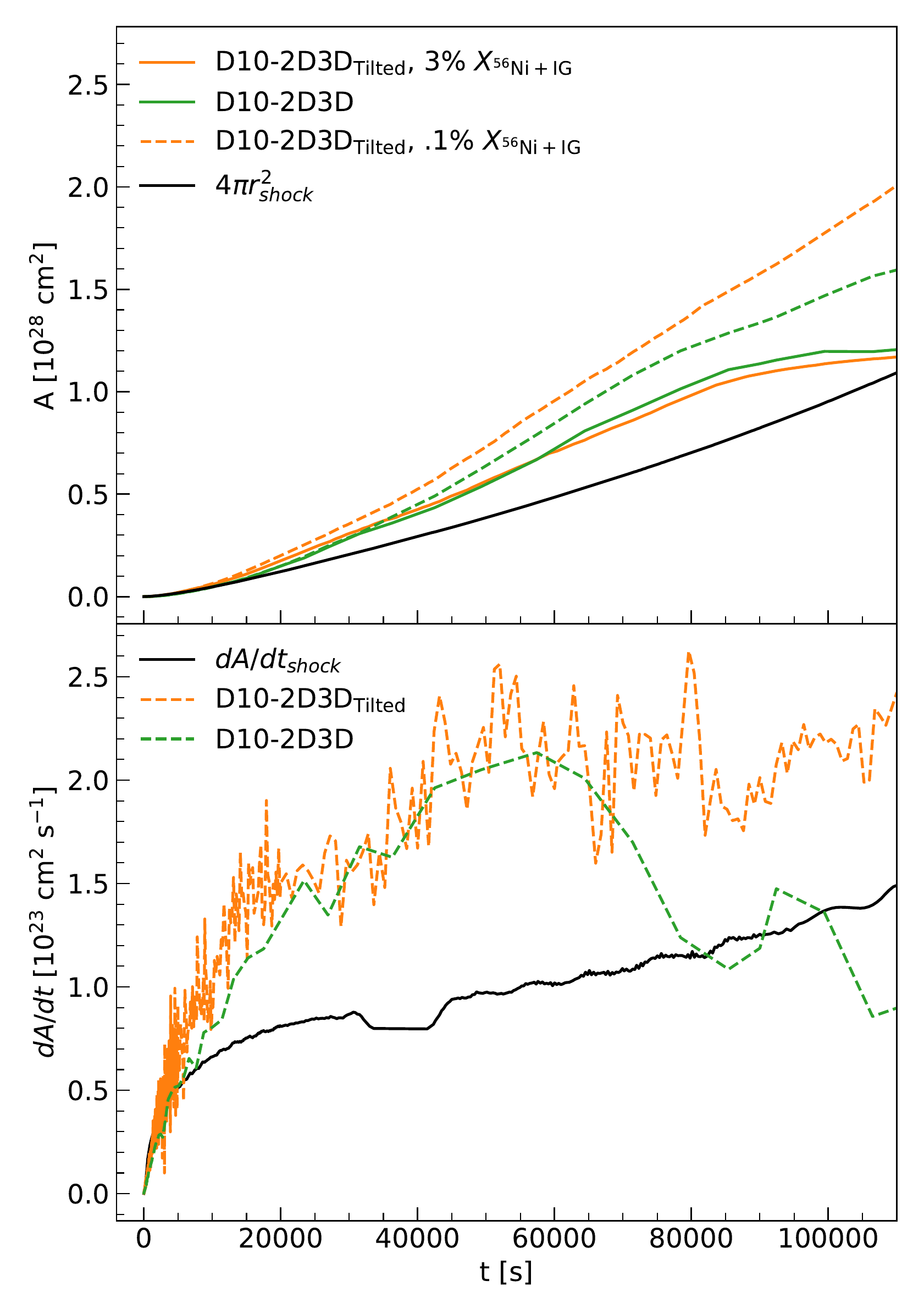}
		\caption{Top: Surface area of the $X_{\rm \isotope{Ni}{56}+IG} = 3\% $ (colored, solid) and $X_{\rm \isotope{Ni}{56}+IG} = 0.1\% $ (colored, dashed) isosurfaces for each D10 model. The average shock radii over time across all models are nearly identical, thus only the surface area of the D10--\spun\ shock (black, solid) is included. Bottom: Numerical time derivatives of the surface area for the shock (black, solid) and $X_{\rm \isotope{Ni}{56}+IG} = 0.1\% $ (colored, dashed) curves of the top plot. Note that the difference in file output in the D10--\spun\ simulation has led to a less dense distribution of data points.} \label{fig:d10area}
	\end{figure}

	The dynamics of the small features are further demonstrated by the growth in the isosurface areas (Figure~\ref{fig:d10area}).
	The total area for both the external coating (0.1\% isosurface) and inner anatomy (3\% isosurface) of the \isotope{Ni}{56}+IG-rich plumes start to diverge early during the dramatic acceleration of the shock. 
	After encountering the reverse shock at $\sim$10000~s, the total surface area represented by the external coating (0.1\% isosurface) of the bullets diverges further, as the bullets in D10--\tilt\ are able to burrow through it more efficiently due to the somewhat higher velocities that result from the spherical-like structures created upon the deviation from axisymmetry.
	The second divergence between models is not present in the 3\% isosurface (inner anatomy) curves.
	This is not surprising due to the relatively similar distribution of metal-rich ejecta in both simulations, with the key differences occurring at larger mass coordinates and higher velocities that are inherently captured by the external coating isosurface instead.
	As with D9.6, the contributions to the total surface area converge back toward those provided by the expansion of the shock and more dramatically for D10--\spun, which stays more axisymmetric and lacks the spherical-bubble structures that retain higher velocities and prolong the convergence to the shock-driven area increase.
	In contrast to the D9.6 models, the 0.1\% isosurface in the D10 models has a larger total surface area than the 3\% isosurface due to the considerable amount of fallback caused by the reverse shocks combined with a more condensed angular distribution of the metal-rich ejecta due to fewer R-T plumes spanning the whole volume.
	
	The D10--\tilt\ model achieves greater velocities compared to D10--\spun, although not as striking as D9.6--\tilt.
	We believe that this is less dramatic in this simulation due to the resolution-limited spawning of only a few clumps at the He/H fragmentation, whereas D9.6 has a wider range of bullets developing from its (C+O)/He fragmentation.
	We would expect a larger deviation from the D10--\spun\ toroid shape if the fragmentation environment was more similar to the high resolution test of Figure~\ref{fig:d102dcomp}.
	Most importantly, despite all this, the end result of the D10--\tilt\ model no longer looks like a rotated 2D model and provides a more faithful 3D model of a polar-dominated explosion than the un-tilted D10--\spun\ model.
	\newpage

	\section{Conclusions} \label{conc}
	We have computed simulations of CCSNe using the FLASH code from the end of the neutrino-driven phase until shock breakout using two stellar progenitors with different structures, a 9.6~\msun\ zero-metallicity RSG and a 10~\msun\ solar-metallicity RSG.
	We have performed these simulations using 160 nuclear species --- the largest network ever used in this regime --- and  higher resolution than comparative studies to provide a more faithful rendering of the composition, development, and terminal distribution of R-T plumes.
	
	The fully-consistent 3D model, D9.6--\threed, develops extended structures out of the He/H interface that are fingerprints of the early asymmetries present in the \chimera\ model. 
	This agrees with the general findings of \citet{WoMuJa15} regarding their analysis of morphology development of different progenitors.
	The density profile of this star allows for steady deceleration of the shock through the He shell, which keeps the leading R-T bullets close to the rear of the shock.
	Consequently, the He/H mass shell has great impact on the trailing ejecta after the shock front has collided with the interface, with the ability to trap the bulk of the metal-rich ejecta if the R-T bullets are moving too slow relative to the shock.
	Because the relative velocity gap between $v_{\rm Shock}$ and $v_{\rm bullets}$ is small enough in D9.6--\threed, the leading R-T bullets (those representing the greatest early-time asymmetries) are not trapped behind the wall of \isotope{He}{4}, reaching velocities of $\sim$1750~\kms.
	
	Our 2D3D D9.6 simulations show that in the absence of a fully-consistent 3D model, tilting the axis of an axisymmetric 2D model in 3D produces a final morphology which better resembles a fully 3D model.
	The rotation of the coordinates breaks the symmetry of the non-radial velocities such that the initially toroidal structure of the 2D-to-3D model develops spherical-bubble structures along its originally axisymmetric toroids (D9.6--\tilt), which does not occur when the 3D grid remains aligned to the original 2D symmetry axis (D9.6--\spun).
	These bubbles retain higher velocities and more easily spawn further R-T plumes at key density interfaces, which directly affects the final morphology of the ejecta.
	Because of the lack of spherical-bubble structures, the leading bullets in the D9.6--\spun\ model move slow enough to get trapped behind the He/H wall (limiting velocities to $\sim$1250~\kms\ at shock breakout), thus this model does not share the morphological development of the D9.6--\threed\ model.
	Therefore, the D9.6--\spun\ model looks primarily like a 2D model that has been extended to 3D space in axisymmetry --- even at shock breakout.
	
	A similar trapping event occurs in the studies of \citet{StJaKr20} for the same progenitor, as their model does not develop distinct elongated structures beyond the He wall, even though it is a fully 3D model.
	The morphological contrast is most clearly seen by comparing Figure 20 in \citet{StJaKr20} with our Figure~\ref{fig:d96velr}.
	The result is a distribution of ejecta in both mass and velocity spaces that looks much more like the distribution seen in our effectively 2D D9.6--\spun\ model.
	We believe this divergence in behavior for similar codes modeling the same progenitor is due to the explosion in the \vertex\ model being much less powerful than that in the \chimera\ model, as the diagnostic explosion energy of our input explosion model is 95\% larger.
	The lower explosion energy of the \vertex\ model does not allow the Ni bullets to retain sufficient velocities to keep up with the shock, leading to the 250\% difference we see at shock breakout between our maximum \isotope{Ni}{56} velocities and theirs.
	This, in combination with our angular resolution being twice as high, leads to different R-T fragmentation developing from the density interfaces.
	
	The D9.6--\tilt\ model develops extended structures beyond the He/H interface, and also maintains maximum velocities of the metal-rich clumps similar to D9.6--\threed.
	This enables further mixing of metal-rich ejecta into the outer regions of the H envelope, thus providing similar ejecta distribution in both mass and velocity spaces, with the bullets reaching $\sim$1950~\kms\ at shock breakout.
	Clearly, the D9.6--\tilt\ model shows that axisymmetry is able to be broken with minimal perturbations.
	
	We applied the same tilting comparison to the D10 progenitor, as we did not have a corresponding 3D \chimera\ model that has achieved a successful explosion.
	We acknowledge that tilting, because of the cutout along the polar axis in the FLASH model, does include more of the polar flow onto the grid, yet emphasize this is extremely dependent on the initial conditions of the 2D model, as the polar flows are particularly strong in the D10 models (as opposed to the D9.6 where polar flows in all models are comparable to flows at other latitudes).
	In mass and velocity spaces, we see relatively consistent distributions in both models, but we still see higher velocities and more outward radial mixing in the D10--\tilt\ model ($\sim$2500~\kms) when compared to D10--\spun\ ($\sim$1900~\kms).
	The parameterized 18~\msun\ and 19.8~\msun\ RSG models of \citet{OnNaFe20}, which have density profiles past the (C+O)/He interface that are similar to our D10 progenitor, achieve even higher velocities ($\sim$5000~\kms), but this is due to a significantly larger explosion energy in their models ($\sim$\ten{1.8}{51}~ergs compared to $\sim$\ten{3.1}{50}~ergs in our model).
	Although D10--\tilt\ did not have as extreme an effect on the distribution of ejecta as was seen in the D9.6 model, tilting seems to have few drawbacks and significant benefits by breaking the toroidal symmetry and restoring a more natural structure to the final distribution of ejecta.
	
	As with the D9.6 models, the D10--\spun\ and D10--\tilt\ models are also consistent with the morphology analysis of \citet{WoMuJa15}.
	The type of morphology seen in the D10 simulations, a few extremely elongated R-T fingers, is due to the strongly varying density profile the shock encounters during its progression.
	The strong acceleration of the shock before encountering the He/H interface creates a large separation between the shock front and metal-rich clumps, thus allowing those metal-rich R-T plumes to grow quite elongated before catching up to the reverse shock created from the subsequent deceleration of the main shock.
	Examining the D9.6 and D10 models, we stress the importance of the density structure on the evolution of the explosion, as widely different results occur depending on the shock progression through the stellar density interfaces.
	However, as we discussed earlier in this section, the contrast between the \citet{StJaKr20} z9.6 model and our D9.6--\threed\ model highlights the ability of the strongly aspherical initial explosion launched by the neutrino-driven, convective central engine to mediate the influence of the progenitor structure.  
	
	We believe the minimal impact of tilting on D10 is due to the strong polar flow and nature of the density profile in this progenitor.
	The more complex system of D9.6--\tilt, with more R-T plumes across all latitudes, is more strongly affected by the tilting.
	In contrast, a simulation with few dominant R-T plumes does not provide enough dynamics between the longitudinal and latitudinal velocities to drive a clear deviation from axisymmetry (D10--\tilt).
	Nevertheless, from a morphological standpoint, the D10--\tilt\ model \emph{still} appears more realistic than D10--\spun.
	The fact that the sole difference between the \spun\ and \tilt\ models is that the initial conditions are rotated 90\degr, and that this causes an originally axisymmetric model to behave more like a 3D model, is a fascinating discovery.
	Although this seems to be progenitor and potentially resolution dependent, this gives much more value to a pure 2D model than previously believed.
	Most importantly, as opposed to more invasive approaches that explicitly inject the system with artificial velocities, tilting the model conserves momentum and all other grid quantities after the coordinate transform.
	Because of the simplicity to extending a 2D model like this in 3D, we recommend exploring this approach if one does not have a true 3D model available.
	However, while the tilting method remedies some of the more egregious problems with un-tilted \spun\ models, we caution that this approach does not fully replicate a 3D model and is by no means a perfect replacement.
	
	Analyzing the distribution of ejecta for both of these progenitors shines light on the importance of using a realistic nuclear network.
	That the total mass yields of our neutron-rich material rivals \isotope{Ni}{56} --- and in some cases exceeds it --- shows the importance of tracking a realistic number of species throughout the entire explosion, not just during the neutrino heating phase.
	This is highlighted by the extent of radial mixing we see of this neutron-rich material into the outer envelope (extending to the surface in both progenitors).
	In our D9.6 simulations, we also see a higher abundance of \isotope{Ni}{60} than \isotope{Ni}{56} in high-velocity regions, $v \gtrsim$~1750~\kms.
	Although others, such as \citet{StJaKr20}, tried tracking neutron-rich material with a tracer nucleus, our results strongly imply that a tracer nucleus does not fully capture the yields or distribution of neutron-rich material at shock breakout, as demonstrated by our yields being an order of magnitude larger and extending significantly passed the $\sim$2~\msun\ and $\sim$500~\kms\ maximum extents seen in \citet{StJaKr20}.
	Of course, the largest difference between z9.6 of \citet{StJaKr20} and our D9.6--\threed\ simulation are the results of the respective \vertex\ and \chimera\ runs.
	The larger explosion energy and larger quantity of heavy element ejecta limits our ability to compare the results of the tracer nucleus approach to our realistic nuclear set. 
	But these differences also act as a reminder that although these extended simulations further develop the final \emph{distribution} of the ejecta, the \emph{amount} of ejecta seen at shock breakout --- and the final fate of the supernova --- is determined by the explosion at early epochs.
	
	\acknowledgments
	 The authors would like to acknowledge valuable conversations with Chloe Sandoval, Steve Bruenn, Rodrigo Fernandez, Thomas Janka, and Ewald M\"uller.
	 This research was supported by the National Science Foundation, Nuclear Theory program under grants PHY-1516197 and PHY-1913531 and by the U.S. Department of Energy, Office of Science, Nuclear Theory program.
	 Additional support was provided by the Exascale Computing Project (17-SC-20-SC), a collaborative effort of the U.S. Department of Energy Office of Science and the National Nuclear Security Administration. 
	 This material is based upon work supported by the U.S. Department of Energy, Office of Science, Office of Advanced Scientific Computing Research and Office of Nuclear Physics, Scientific Discovery through Advanced Computing (SciDAC) program.
	 Research at Oak Ridge National Laboratory is supported under contract DE-AC05-00OR22725 from the U.S. Department of Energy to UT-Battelle, LLC.  
	 An award of computer time was provided by the Innovative and Novel Computational Impact on Theory and Experiment (INCITE) program. 
	 This research used resources of the Oak Ridge Leadership Computing Facility, which is a DOE Office of Science User Facility supported under Contract DE-AC05-00OR22725.
	\facilities{OLCF}
	\software{FLASH \citep{FrOlRi00,DuAnGa09}, NumPy \citep{NumPy20}, SciPy \citep{SciPy20}, Matplotlib \citep{Matplotlib07}, h5py \citep{h5py13}, Blender (\url{www.blender.org}), VisIt \citep{VisIt12}}
	
	\bibliographystyle{aasjournal}
	
\end{document}